\documentclass[hyper,a4paper,12pt]{JHEP}
\usepackage{graphicx}
\usepackage{amsmath,epsfig,amssymb,multirow}
\allowdisplaybreaks
\usepackage{slashbox}
\newcommand{\fatone}{1\hspace{-0.8ex}1\hspace{0ex}}

\newcommand{\tr}{\mbox{tr}}
\newcommand{\sub}[2]{{#1}_{\mbox{\tiny{#2}}}}
\title{An Exceptional SSM from $E_6$ Orbifold GUTs with intermediate LR symmetry}
\author{Felix Braam\footnote{email:
    felix.braam@physik.uni-freiburg.de}, 
  Alexander Knochel\footnote{email:
    alexander.knochel@physik.uni-freiburg.de}, 
  J\"urgen Reuter\footnote{email: reuter@physik.uni-freiburg.de}\\
Albert-Ludwigs-Universit\"at Freiburg \\
Physikalisches Institut\\
Hermann-Herder-Str. 3 \\
79104 Freiburg, GERMANY}
\abstract{We propose a class of $E_6$-based local orbifold Grand
  Unified Theories (GUTs) which
  yield an exceptional supersymmetric standard model as their low
  energy theory including leptoquark and un-Higgs exotics and a $Z'$ at
  the TeV scale. Unification is achieved in two steps through an
  intermediate scale symmetry breaking at a unified coupling which 
is enhanced due to the contributions of leptoquarks to the QCD beta function.}    
\preprint{FR-PHENO-2010-08, 1001.4074 [hep-ph]}
\begin{document}
\section{Introduction}

The attractiveness of supersymmetry in contemporary particle physics
model building is based to some extent on the observation that in the
minimal (MSSM) and next-to-minimal supersymmetric standard model
(NMSSM) the coupling constants of the electroweak and strong
interactions unify around $10^{16}$ GeV. Beyond that scale, all
interactions apart from gravity might be unified into one fundamental
force, mathematically modeled by a simple gauge group. Unfortunately
there is no simple gauge group providing a set of irreducible
representations which exclusively contain the (N)MSSM particle 
content. Therefore one has to either argue why the additional
(``exotic'') particles completing the GUT representations become
heavy (``doublet-triplet splitting'') \cite{Sakai:1981gr}, or take
into account their effect on the running of the gauge couplings if
they remain in the spectrum below the unification  scale.

In the latter approach, the gauge couplings do not unify as in the (N)MSSM.
Unification can be restored by introducing light incomplete GUT 
representations \cite{Athron:2009is} or postulating an intermediate
gauge symmetry\footnote{The breaking scale of this symmetry could
ideally be linked to a Majorana mass-term for the right-handed
neutrino, triggering a see-saw mechanism and hence naturally
generating small neutrino masses.}. In \cite{Kilian:2006hh} it has
been pointed out that unification of the gauge couplings can be
achieved even in the presence of complete $\mathbf{27}\; E_6$ matter if
the interactions above $\Lambda_{\mbox{\tiny{int}}}\sim10^{15}$ can be
described by an $G_{PS} \times U(1)_\chi \equiv SU(4)\times SU(2)^2 
\times U(1)_\chi$ gauge symmetry.   As successful this approach is
in the gauge coupling sector, it fails to yield a viable low energy
theory:

Since in each generation, the $E_6$ scenario unifies matter, Higgs fields and
exotics in the fundamental $\bf{ 27}$ representation, there is only one
singlet in the renormalizable superpotential,
\begin{equation}
  \label{we6}
  \mathcal  W \sim \bf 27\,27\,27
\end{equation}
which does not discriminate between NMSSM type terms 
  \begin{equation}
       \mathcal{W}_{NMSSM}\sim SHH+ S D^c D + H L l^c +H Q q^c
  \end{equation} 
  and leptoquark- and diquark-like couplings 
  \begin{equation} 
       \mathcal{W}_{LDQ} \sim D q^c l^c + D^c Q L+ D Q Q + D^c q^cq^c
  \end{equation}
  which lead to fast proton decay if the triplet Higgs
  like  exotics $D_i$, $D^c_i$ have masses significantly below the GUT scale.
  Furthermore, as there are three copies of an NMSSM-like Higgs sector, the 
  postulate of an H-Parity \cite{Griest:1990vh} allowing only one Higgs 
  generation to couple to matter is very effective in explaining the absence 
  of large flavor changing neutral currents (FCNCs), but inconsistent with the 
  $E_6$-symmetric superpotential (\ref{we6}).

If $E_6$ is broken to $SU(4)\times SU(2)^2 \times U(1)_\chi$ close to the 
Planck scale, the lepto/diquark couplings become a separate singlet
and can in principle be strongly suppressed in the low energy
theory. This could be achieved if the renormalizable superpotential
vanished in the high energy theory and is only generated from higher
dimensional operators in the course of the $E_6$ breaking
\cite{spurionpaper}, or if matter resides on a symmetry-reduced fixed
point of an orbifold. In the case of intermediate Pati-Salam symmetry
which unifies quarks and leptons in $\bf {4}$ and $\bf{ \overline 4}$  
representations, the intermediate breaking scale is generically several 
orders of magnitude below the Planck scale \cite{Braam:2009fi},
\begin{equation}
  \Lambda_{PS}\ll \Lambda_{E6}\sim \Lambda_{Pl}
\end{equation}
predicting that the exotics $D$, $D^c$ will neither have large leptoquark- nor
diquark-like couplings. Thus, they behave like heavy, relatively long lived
$R$-odd right-handed quarks in the low energy theory. This can only be relaxed
if the intermediate theory does not leave $SU(4)$ intact too far below the
Planck scale, which makes it more unnatural to associate $\sub{\Lambda}{int}$ 
with the see-saw scale.

\begin{figure}
     \begin{center}
       \includegraphics[width=0.48\linewidth]{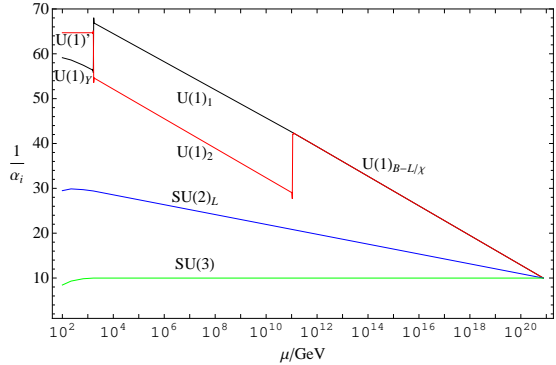}
     \end{center}
     \caption{The unification scenario featuring three full $\bf 27$
       multiplets of $E_6$ and an intermediate $G_{LR}\times U(1)_\chi$
       symmetry which is broken at the scale $\sub{\Lambda}{int}$. The
       gauge couplings do not unify to $E_6$ below the Planck scale.}  
     \label{fig:LRwo27}
   \end{figure}

The latter objections could be resolved by reducing the degree of symmetry 
in the intermediate regime to a minimal left-right symmetric (N)MSSM 
 with 
\begin{equation} G_{LR}\times U(1)_\chi =SU(3)_C\times SU(2)_L \times
  SU(2)_R \times U(1)_{B-L} \times U(1)_\chi \end{equation}  
gauge symmetry. However, as illustrated in Fig.~\ref{fig:LRwo27}, in
that case the coupling constants unify only far above 
the Planck scale. 

The problems mentioned above seem to rule out simple $E_6$-invariant GUT
set-ups in the absence of doublet-triplet splitting in the spectrum, and therefore lead us to consider higher
dimensional orbifold constructions. Orbifold breaking has been
proposed in \cite{Dixon:1985jw,Dixon:1986jc,Bailin:1987pf} for string
models. The breaking of exceptional groups in extra dimensions has
been of considerable interest in recent years in the context of
stringy constructions \cite{Kobayashi:2004ud, Kobayashi:2004ya,
  Buchmuller:2006ik, Lebedev:2006kn,
  Lebedev:2006tr,Lebedev:2007hv,Nibbelink:2009sp} as well as field
theoretic models \cite{Asaka:2001eh,Asaka:2002nd,Hebecker:2003jt,
  Forste:2005rs}, most of which aim to reproduce an MSSM-like low
energy theory to achieve standard unification. 

From the point of view of quantum field theory, the low (or
intermediate) scale gauge symmetry which we want to realize is reached in the
local GUT ansatz by projecting out parts of the $E_6$ roots by point
transformations in the extra dimensions. The corresponding
gauge bosons become massive by a geometric Higgs mechanism.
The fixed points of the orbifold support reduced 4D gauge theories,
corresponding smaller representations and more general
superpotentials, while the bulk of the extra dimensional volume
retains the full gauge invariance. This construction ameliorates the
severe constraints on the low-energy theory and allows us to obtain an
Exceptional SSM scenario in which the effective $\mu$ term is
generated by an MSSM singlet vev $\langle S \rangle $ as in the NMSSM. Furthermore, our
modified unification scheme allows the full $E_6$ matter spectrum in
the 4D theory and thus naturally yields an anomaly-free extra $U(1)$
factor. This generates the abovementioned singlet vev via its $D$
term, thus circumventing the cosmological problems of domain walls
which can plague NMSSM models with an $S^3$ superpotential term. 


\section{An Exceptional SSM with intermediate PS Symmetry}
\label{sect:5dmodel}
Before we turn to exceptional unification based on a $G_{LR}\times U(1)_\chi$ intermediate symmetry
which requires orbifold geometries in $D=6$ to achieve the necessary degree of symmetry breaking, 
we discuss possibilies and limitations of models with a $PS \times U(1)_\chi$ intermediate symmetry from 5D orbifolds.

\subsection{$E_6\rightarrow PS \times U(1)$ Breaking on $S^1/(\mathbb Z_2 \times \mathbb Z_2')$}
Technical details about the $S^1/\mathbb Z_2$ orbifold construction can be found in Appendix \ref{sec:5dgeometries}.
The orbifold reflections $r$ and $r'$ can be endowed with order two gauge shifts to break $E_6$ to local $SO(10)\times U(1)$ or $SU(6)\times
SU(2)$ invariance on the fixed points.  We can thus break $E_6$ down to $PS \times U(1)_\chi$ 
by breaking to $SU(6) \times  SU(2)_L$ and $SU(6) \times SU(2)_R$, or, more interesting for our purpose, 
to $SU(6)\times SU(2)_{L/R}$ and $SO(10)\times U(1)_\chi$ with $\mathbb Z_2$ and $\mathbb Z_2'$ respectively.
If matter is localized on a $SO(10)\times U(1)_\chi$ fixed point, we have naturally
light Higgs triplets in the $\mathbf{10}$, and we obtain leptoquark-
and diquark-like couplings of matter to the Higgs sector from
the $\mathbf{10\; 16\; 16}$ superpotential term. Similarly in the $SU(6)$ case, the exotics
are in the $\mathbf{\overline{15}}$ together with left- or right-handed matter and the 
singlet, and lepto- and diquark couplings are contained in both superpotential
terms $\mathbf{6 \; 6\; \overline{15}}$ and $\mathbf{
  \overline{15}\,\overline{15}\,\overline{15}}$. This means that in this scenario it is not possible to realize exceptional unification
with light exotics using brane localized matter, because lepto-diquark interactions are not suppressed. However,
one can attempt to put the third generation which contains the MSSM Higgs doublet, the NMSSM-like singlet and color charged exotics, in the 5D Bulk to avoid
lepto-diquark couplings while allowing quark masses at the same time. This would be a kind of orbifold-based doublet-triplet splitting, yet with what looks like complete $E_6$ multiplets below the compactification scale enforced by anomaly constraints. The $SU(6)\times SU(2)_{L/R}$ fixed point should
only contain vectorlike matter such as the Higgs sector connected to the intermediate breaking.
We now consider the $SU(6)\times SU(2)_L$ case, the case $SU(6)\times SU(2)_R$ is completely analogous
except for the form of the allowed brane superpotentials.
\begin{figure}
\begin{center}
\includegraphics[width=12cm]{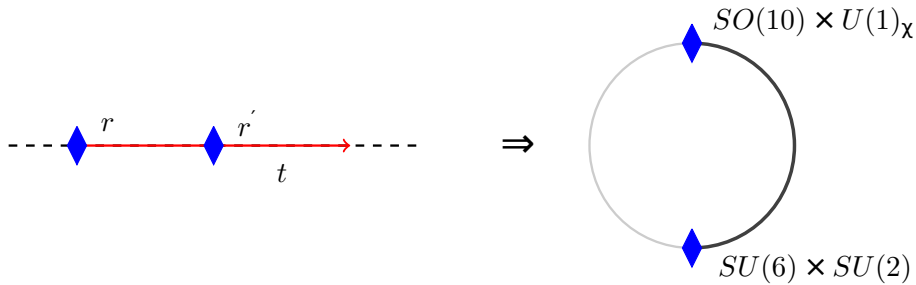}
\end{center}
\caption{An $S^1/(\mathbb Z_2 \times \mathbb Z_2')$ orbifold breaking of $E_6 \rightarrow PS \times U(1)_\chi$. The two order 2 reflections are shown as blue diamonds,
the translation which compactifies $\mathbb R \rightarrow S^1$ is shown as a red arrow.\label{fig:s1orbifold}}
\end{figure}

\subsection{Brane and Bulk Matter} Chiral components of the 5D gauge
hypermultiplets and matter hypermultiplets can induce brane localized anomalies 
even though they are not in the massless spectrum. 
The 5D gauge multiplet can be written as a vector-
and a chiral superfield and thus contains the component fields $\hat V\ni
A_\mu, \lambda_1$ and $\hat \chi\ni \Sigma,\lambda_2$ where $\lambda_i$ are 4D
Weyl spinors and $\Sigma$ is a complex scalar containing $A_5$ and an
additional real scalar. In a $\mathbb Z_2$ orbifold, each 4D gauge superfield
$\hat V^\alpha$ corresponding to a root $\alpha$ which is projected out entails
a chiral superfield $\hat \chi^\alpha$ which survives on this fixed point.
Since the adjoint of $E_6$ and of the unbroken subgroup are vectorlike
representations, the surviving chiral modes also come in real representations,
namely $\bf 16_{-3/2}+\overline{16}_{3/2}$ on the $SO(10)\times U(1)_\chi$
fixed point, and $\bf (20,2)$ on the $SU(6)\times SU(2)$ fixed point. The 16
chiral zero modes which survive on both fixed points therefore correspond to a
vectorlike representation as well. 

Now, let us consider whether one can obtain full anomaly free generations from
$\bf 27$ representations in the bulk. 5D matter hypermultiplets consist of two
oppositely charged chiral superfields  $\hat\Phi\ni \psi, \phi$ and $ \hat\Phi_c\ni
\psi_c, \phi_c$, which together contain a 5D dirac fermion and corresponding
scalar partners. In a $\mathbb Z_2$ orbifold, the two receive opposite boundary
conditions, meaning that if $\hat \Phi$ is projected out on a fixed point, $\hat\Phi_c$
survives and vice versa. The chiral part of the hypermultiplet transforms under
$\mathbb Z_2$ and $\mathbb Z_2'$ projections as 
\begin{eqnarray}
\hat\Phi^\mu& \stackrel{\mathbb Z_2\phantom{'}}{\longrightarrow}&\sigma \exp[2 \pi i V\phantom{'}\cdot \mu] \hat\Phi^\mu \nonumber \\
\hat \Phi^\mu& \stackrel{\mathbb Z_2'}{\longrightarrow}&\sigma'\exp[2 \pi i V'\cdot \mu] \hat\Phi^\mu
\end{eqnarray}
if $\mu$ is the $E_6$ weight of the hypermultiplet, $V$ and $V'$ are the
gauge shifts associated with $\mathbb Z_2$ and $\mathbb Z_2'$ and the parities
are $\sigma, \sigma' \in \{+,-\}$.  The antichiral part $\hat{\overline{\Phi}}_c$ has opposite parities.
To obtain a complete light $\bf 27$
from bulk hypermultiplets, one needs four $\bf 27$s with parities
$(\sigma,\sigma')=(+,+),(-,-),(-,+),(+,-)$, yielding chiral zero modes $\bf
(6,1,1)_{-1}+(1,1,1)_2$, $\bf (4,2,1)_{\frac{1}{2}}$,
$\bf (\overline{4},1,2)_{\frac{1}{2}}$ and $\bf (1,2,2)_{-1}$ respectively with our choice of gauge shifts, 
\begin{eqnarray} 
\overline V=(\frac{1}{2},\frac{1}{2},0,\frac{1}{2},\frac{1}{2},0),\quad  \overline V'=(\frac{1}{2},\frac{1}{2},\frac{1}{2},\frac{1}{2},\frac{1}{2},0).
\end{eqnarray}
However, the corresponding zero modes from the antichiral multiplets also form
a complete $\bf 27$, such that the complete matter content is vectorlike. 

The possibility remains to obtain part of a $\bf 27$ generation, such as the Higgs sector, from the
bulk, while the remaining chiral components are localized on the $SO(10)\times
U(1)_\chi$ fixed point. 
We are now going to implement what corresponds to a $\bf 10_{-1} +1_2\in 27$ as the zero modes of two bulk hypermultiplets in the $\bf 27$.
We call the two hypermultiplets $\bf 27^A$ and $\bf 27^B$ and give them $\mathbb Z_2$ parities $(+,+)^{(A)},(+,-)^{(B)}$. This produces
massless chiral multiplets corresponding to $$\bf (6,1,1)^A_{-1}+ (1,1,1)^A_2 + (1,2,2)^B_{-1} \sim 10_{-1} +1_2.$$ 
The superscript denotes the $\bf 27$ from which the representations originate.
In particular, these fields contribute the MSSM Higgs doublet and the NMSSM singlet accompanied by a generation of color charged exotics.
The antichiral parts of the bulk hypermultiplets which we now call $\bf 27^{Ac}$ and $\bf 27^{Bc}$, also contain zero modes.
$\bf 27^{Ac}$ contributes extra $SU(2)_L$ charged matter, while $\bf 27^{Bc}$ contributes extra $SU(2)_R$ charged matter, corresponding to representations
$$\bf (4,2,1)^{Ac}_{\frac{1}{2}} + (\overline{4},1,2)^{Bc}_{\frac{1}{2}} \sim 16_{\frac{1}{2}}^c, $$
however with opposite 4D chirality compared to standard matter.
The modes surviving on the $SU(6)\times SU(2)$ fixed point come with opposite chirality from the two hypermultiplets, 
and are thus anomaly free. The modes surviving on the $SO(10)\times U(1)_\chi$ fixed point are $\bf 10_{-1}^A, 1_2^A$ and $\bf 10_{-1}^B, 1_2^B$  
in the chiral part and $\bf 16_{\frac{1}{2}}^{Ac} +
16_{\frac{1}{2}}^{Bc}$ in the antichiral part. To cancel the 4D
anomaly and to complete the matter content of the generation, we need
brane localized chiral $\bf
16_{\frac{1}{2}}^\prime+16_{\frac{1}{2}}^3$, where one linear
combination gets a mass with the antichiral bulk zero mode $\bf
16_{\frac{1}{2}}^{c}$, and the other remains in the light spectrum,
e.g. as the third generation matter. The brane anomaly on the
$SO(10)\times U(1)_\chi$ fixed point, written in terms of chiral
fields, now receives contributions from the two localized chiral modes
$\bf 16^\prime_{\frac{1}{2}}+16^3_{\frac{1}{2}}$ and bulk contributions
from  
${2 \times} {\bf 10_{-1}}+{2 \times} \bf 1_2$ and $2 \times \bf \overline{16}_{-\frac{1}{2}}$. These contributions cancel since the bulk modes only contribute half of the anomaly on each of the two branes.

Such a general setup could be used to implement the ESSM models with Pati-Salam intermediate symmetries \cite{Athron:2009is,Kilian:2006hh}. Note that the couplings of the light exotics to matter, which unify lepto- and diquark terms due to the intermediate Pati-Salam symmetry,  must be suppressed to avoid rapid proton decay, but should not be exactly zero as this would render the color charged exotics stable. 
\subsection{Superpotential}
The abovementioned setup allows in principle to introduce all superpotential
terms necessary for the ESSM on the branes while avoiding lepto-diquark
couplings and unhiggs-matter couplings. Let $\bf 16^i, 10^i,1^i$ with $i=1,2$
be the first two generations of matter, unhiggs and unsinglet fields localized as chiral multiplets on the $SO(10)\times U(1)_\chi$ fixed point. The third generation is implemented as explained in the last section, and uppercase superscripts $\bf A,B$ denote the two bulk hypermultiplets. Then, the following brane localized terms can be introduced.
Matter receives masses from $\bf \langle10^B \rangle  16^a 16^b $ where
$a,b=1..3$. Color charged and unhiggs exotics receive masses from $\bf \langle
1^A \rangle 10^a 10^b$ where $a,b=1,2,A$. Unsinglet-unhiggs mixing is generated
by $\bf 1^j  10^i \langle 10^B  \rangle $ where $i,j=1,2$. The $\mu$ term
contribution can arise on both branes, namely through $\bf \langle 1^A \rangle 10^B 10^B$ on
the $SO(10)\times U(1)_\chi$ fixed point, and through $\bf \langle (\overline{15},1)^A\rangle
(6,2)^B (6,2)^B$ on the $SU(6)\times SU(2)$ fixed point. 
Lepto-diquark interactions are contained in a separate singlet, $\bf 10^A 16^a 16^b$ where $a,b,=1..3$. While the corresponding coupling can assumed to be small, it would be interesting to investigate how its suppression can be enforced, and which bounds result for proton and
exotic decay lifetimes. 

This type of construction
seems to exhaust the possibilities in $D=5$ apart from introducing additional breaking from brane or bulk vevs,
and higher dimensional operators. In particular, in order to obtain
smaller intermediate rank 6 gauge symmetry such as $G_{LR} \times U(1)_\chi$,
or fixed points with smaller gauge groups
which for example allow for light leptoquark exotics and more freedom in the 
matter yukawa couplings, we need to study $E_6$
breaking in higher dimensions.  
\begin{table}
  \begin{center}
    \begin{tabular}{|c||c|c|c|} \hline
      \backslashbox{\footnotesize
      \raisebox{1ex}{$SU(6)\times SU(2)_L$}}{\footnotesize
      \raisebox{-1ex}{$SO(10)_{Q\chi}$}} & $\bf 16_\frac{1}{2}$ & $\bf
      10_{-1} $ & $\bf 1_2 $ \\ \hline \hline 
      $\bf (\overline{15},1)$ & $\bf (\overline 4,1,2)_{\frac{1}{2}} $ & $\bf (6,1,1)_{-1} $ & $\bf(1,1,1)_2$ \\ \hline
      $\bf (6,2)$ & $\bf (4,2,1)_{\frac{1}{2}} $ & $\bf (1,2,2)_{-1}$ & \large $\times$ \\ \hline
\end{tabular}
\end{center}
  \caption{The decomposition of a $\mathbf{27}$ of $E_6$ for the
    rank 6 subgroups $SO(10)\times U(1)_\chi$, $SU(6)\times SU(2)_L $ and their intersection $PS\times U(1)_\chi =
    SU(4)_C\times SU(2)_L\times SU(2)_R \times
    U(1)_\chi$ which is shown inside the table. The subscript denotes
    $\sqrt{6}Q_\chi$, where the
    $U(1)$ generator is $SU(N)$ normalized. The case with $SU(6)\times SU(2)_R$ is completely analogous.\label{tab:decomp5}} 
\end{table}

\section{An Exceptional SSM with intermediate LR Symmetry}

\subsection{$E_6 \rightarrow G_{LR}\times U(1)$ breaking on $T^2/\mathbb Z_6$ \label{sect:model632}} 
\begin{figure}
\begin{center}
\includegraphics[width=7cm]{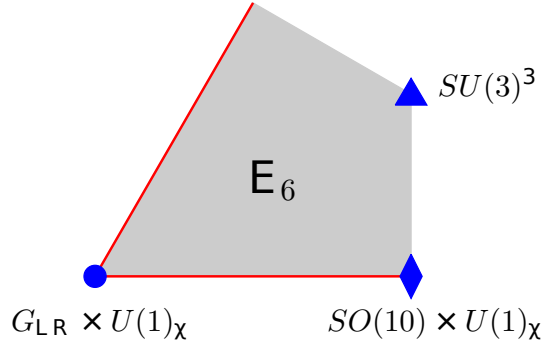}
\end{center}
\caption{An $E_6\rightarrow G_{LR}\times U(1)_\chi$ breaking scenario on the $\mathbb R^2/\bf 632$ orbifold and the local gauge groups at the $\mathbb Z_6$, $\mathbb Z_3$ and $\mathbb Z_2$ fixed points which are shown as blue circle, triangle and diamond. The shaded area shown is the fundamental domain only, and long (red) and short sides are identified. The construction is outlined in Section \ref{sect:geometry632}.\label{fig:fullorbi632}}
\end{figure}

 We now want to turn to
intermediate $G_{LR}$ symmetry including the possibility to assign $H$ parities
and still have leptoquarks at low energies. To reach this goal purely by
orbifold breaking, we require a 6D setup. The technicalities of the 
various relevant constructions based
on torus compactifications are summarized in the Appendix
\ref{sec:6dgeometries}. The following model is based on a
$T^2/\mathbb Z_6$ orbifold in 6D with $E_6$ gauge invariance in the bulk.
$\mathbb Z_6$ geometries have been considered in the context of
$E_8\nolinebreak\times\nolinebreak E_8$ heterotic string compactifications. In this sense, 6D $E_6$ orbifold models can be thought of as an intermediate stage 
from the full string theory construction to a realistic 4D model.
The orbifold geometry and possible phases are
specified in Section \ref{sect:geometry632}. 
The gauge embeddings $G(r_3)$, $G(r_2)$ can be chosen independently, and the induced $\mathbb Z_6$ gauge shift is
$G(r_6)=G(r_3)^{-1}G(r_2)$.  
This means that we can assign shift embeddings $G(r_2): E_6
\rightarrow SO(10)\times U(1)_\chi$, and $G(r_3): E_6 \rightarrow
SU(3)^3$, which will result in $G(r_6): E_6 \rightarrow G_{LR}\times
U(1)_\chi$. This is for example achieved with the shift vector 
\begin{eqnarray}
\label{eqn:shift632}
\overline V(r_6) & = &(\frac{1}{6},-\frac{1}{6},-\frac{1}{3},-\frac{1}{2},-\frac{1}{6},0)
\end{eqnarray}
This is equivalent to a shift in the $\overline Q_{B-L}$ direction, appropriately normalized
to be compatible with the $\mathbb Z_6$ algebra. 
No discrete Wilson line is allowed. This gives us three inequivalent
fixed points with localized gauge invariances $G_{LR}\times U(1)_\chi$,
$SO(10)\times U(1)_\chi$ and trinification $SU(3)^3$.
This can be understood from Table \ref{tab78} since those groups correspond to roots with integer, one-third integer and half integer charge under $Q_{B-L}$, respectively. The construction is shown in Figure \ref{fig:fullorbi632}.
\subsection{Bulk Matter and Anomalies on the $\mathbb R^2/{\bf 632}$ orbifold}
In the SUSY version of the model, vectors are simply extended by massless Weyl gaugino modes in the adjoint, and thus give
us standard 4D $\mathcal N=1$ multiplets corresponding to the unbroken gauge
group. However, the scalar zero modes usually do not appear in vectorlike
pairs and therefore result in massless 
chiral multiplets in complex representations satisfying $V \cdot \alpha \in \mathbb Z + 1/6 $ where V is the $\mathbb Z_6$ gauge shift.
In the case of $E_6\rightarrow G_{LR} \times U(1)_\chi$ breaking on the $\mathbb R^2/{\bf 632}$ orbifold, they contain
quark-like exotics with $Q_\chi= \pm 3/2$, using the notation from Table
\ref{tab78} they are $Q_{16}$ from ${\bf \overline{16}}_{3/2}$ and $U_{16},
D_{16}$ from ${\bf 16}_{-3/2}$ or vice versa, depending on the choice of the
gauge shift.  Note that they do not form complete $SU(3)^3$ or $SO(10)\times
U(1)_\chi$
representations but appear in $G_{LR}$ representations, nor are they by themselves anomaly free with respect to the unbroken gauge group $G_{LR}\times U(1)_\chi$. Since we preserve our extra $U(1)_\chi$ or $U(1)'$ at low energies,
these exotics do not mix with standard matter from the $\bf 27$.  However, they
induce 4D anomalies and localized anomalies on the $\mathbb Z_6$ and $\mathbb
Z_3$ fixed points, where the chiral bulk fermions
from the gauge superfield $\hat\chi$ which would survive the individual orbifold projections on one fixed point are in $\bf (3,2,1)+(3,1,2)$ and $\bf (3,3,\overline{3})$ representations of $G_{LR} \times U(1)_\chi$ and $SU(3)^3$ respectively, while on the $\mathbb Z_2$ fixed point,
vectorlike contributions corresponding to ${\bf 16}_{-3/2}+{\bf \overline{16}}_{3/2}$ of $SO(10)\times
U(1)_\chi$ survive the projection and cancel. To cancel the 4D anomalies, one can put the
completion consisting of a $\bf(\overline 3,2,1)+(\overline 3,1,2)$ on the
$\mathbb Z_6$ fixed point, which leaves open the
problem how the 6D brane localized anomalies are cancelled. A simpler way to cancel the 4D and
brane anomalies while removing the chiral bulk modes is the doubling of the
bulk fermion content with an additional hypermultiplet in the adjoint $\bf 78$.
The hypermultiplet has opposite 6D chirality compared to the 6D gauginos, so the fermions of both form an unconstrained 6D spinor.
The  $\mathbb Z_6$ parity of the hypermultiplet must be chosen such that the surviving 4D chiral modes from $\hat \Phi$ on
each fixed point are in the same representation as the 4D antichiral zero modes from $\hat {\overline\chi}$.
Under a $\mathbb Z_6$ rotation, the vector hypermultiplet belonging to a root $\alpha$ transforms as
\begin{eqnarray}
\hat V^\alpha & \longrightarrow & e^{2 \pi i V \cdot \alpha} \hat V^\alpha \nonumber \\
\hat \chi^{-\alpha} & \longrightarrow & e^{-2 \pi i V \cdot \alpha} e^{- \frac{2\pi i }{6}}\hat \chi^{-\alpha} .
\end{eqnarray}
The chiral superfields in an adjoint hypermultiplet  belonging to a root $\alpha$ transform as
\begin{eqnarray}
\label{eqn:bulkadjoint}
\hat \Phi^{\alpha} & \longrightarrow & e^{2 \pi i V \cdot \alpha}e^{-\frac{2 \pi i}{6}a_+} \hat \Phi^\alpha  \nonumber \\
\hat \Phi_c^{-\alpha} & \longrightarrow & e^{-2 \pi i V \cdot \alpha}e^{- \frac{2 \pi i}{6}a_-} \hat \Phi_c^{-\alpha}
\end{eqnarray}
From invariance of the 6D hypermultiplet kinetic term \cite{HEIDISUSY}, $\int \! d^2\theta\,\hat \Phi_c^{-\alpha} \partial \hat \Phi^\alpha$ (where $\partial=(\partial_5 -i \partial_6)/\sqrt{2}$ transforms like $\partial\rightarrow e^{-2 \pi i/6}\partial$ under orbifold rotations in analogy to the 6D gauge field $\hat \chi$),
$a_+ + a_- \in 6 \mathbb Z-1$.
For the choice $a_+=5$, $a_-=0$, every invariant mode $\hat \chi^{-\alpha}$ is accompanied by an invariant mode $\hat \Phi^{\alpha}$, and the chiral
zero modes from $\hat\chi$ and $\hat \Phi$ appear in mutually complex conjugate representations. The invariant modes of $\hat \Phi_c$ are in the adjoint of the unbroken gauge group, and localized
and 4D gauge anomalies are cancelled and we assume all extra bulk matter to have
masses at the compactification scale.  This has the interesting consequence
that the Kaluza-Klein excitations in the adjoint appear in 4D $\mathcal N=4$ multiplets and
therefore the contributions to powerlike threshold corrections to gauge unification can vanish
\cite{Hebecker:2002vm}.
In a full embedding in $E_8 \times E_8$
heterotic models, after breaking of $E_8^{vis} \rightarrow G_{LR} \times U(1)^3$, all but one of the $U(1)$ factors are automatically anomaly free.
It remains a subject of further study how this could be realized in the context of the exceptional SSM proposed in this paper.

Analogous to the 5D model with intermediate $PS\times U(1)_\chi$ symmetry, one
can consider implementing parts of a generation, for example the Higgs fields,
as 6D bulk fields rather than as being purely localized on a fixed point. The
orbifold transformation properties for a bulk hypermultiplet of weight $\mu$
are analogous to (\ref{eqn:bulkadjoint}),
\begin{eqnarray}
\label{eqn:bulkfundamental}
\hat \Phi^{\mu\phantom{-}} & \longrightarrow & e^{2 \pi i V \cdot \mu}e^{-\frac{2 \pi i}{6}a_+} \hat \Phi^\mu  \nonumber \\
\hat \Phi_c^{-\mu} & \longrightarrow & e^{-2 \pi i V \cdot \mu}e^{- \frac{2 \pi i}{6}a_-} \hat \Phi_c^{-\mu}
\end{eqnarray}
where again, the integers $a_+= 0 \dots 5$ and $a_- = 5-a_+$ can be chosen for each hypermultiplet. 
A complete $\bf 10_{-1}+1_2$ from the bulk could be realized by three $\bf
27$ with parities $a_+ = 0, 2, 4$. With it comes a complete massless $\bf \overline{16}_{-\frac{1}{2}}$
from the antichiral part. However, the resulting brane anomaly structure is
nontrivial \cite{vonGersdorff:2003dt,Scrucca:2004jn}. This construction results in a vanishing
anomaly on the $\mathbb Z_3$ fixed point, and the modes contributing to the
$\mathbb Z_2$ and $\mathbb Z_6$ brane anomalies come in complete $SO(10)$
multiplets. In the conventions given in \cite{vonGersdorff:2003dt,Scrucca:2004jn}, the representations producing the localized anomalies, both equivalent to $({\bf 10_{-1}+1_2 + \overline{16}_{-\frac{1}{2}}})$ on the $\mathbb Z_6$ fixed point and
on the $\mathbb Z_2$ fixed point, contribute with prefactors $\frac{1}{4}$ and $\frac{3}{4}$ respectively to the remaining 4D anomaly. Since the factors are not (half) integers, even
after brane localized matter $2 \times \bf 16_{\frac{1}{2}}$ is added to obtain the correct spectrum and cancel the 4D anomaly, localized $U(1)$ anomalies remain which would have to be cancelled by some mechanism. For the remainder of our discussion we therefore assume brane localized representations corresponding to complete $\bf 27$ to provide the matter content.

\subsection{Local Matter content and gauge unification}

In accordance with the local GUT framework, the massless spectrum on orbifold 
fixed points is not determined by the unbroken 4D gauge invariance, but --
as it is necessary to obtain a consistent gauge theory -- by the local
gauge invariance on the fixed point itself, which is in general larger. 
This is in contrast to bulk matter the massless modes of which generally 
only respect the low energy gauge symmetry.
The branching rules of the anomaly-free field content of a complete
$\mathbf{27}$ of $E_6$ for the rank 6 subgroups $SO(10)\times U(1)_\chi$,
$SU(3)^3$ and their intersection $G_{LR}\times U(1)_\chi$ are given
in Table~\ref{tab:decomp}.  

\begin{table}
  \begin{center}
    \begin{tabular}{|c||c|c|c|} \hline
      \backslashbox{\footnotesize
      \raisebox{1ex}{$SU(3)^3$}}{\footnotesize
      \raisebox{-1ex}{$SO(10)_{Q\chi}$}} & $\bf 16_\frac{1}{2}$ & $\bf
      10_{-1} $ & $\bf 1_2 $ \\ \hline \hline 
      $\bf A\, =\,(\overline 3, 1, 3)  $ 	      & $\bf
      (\overline 3,1,2)_{(-\frac{1}{3},\phantom{-}\frac{1}{2})} $   &
      $\bf (\overline 3,1,1)_{(\phantom{-}\frac{2}{3},-1)} $   &\Large
      $\times$ \\ \hline 
      $\bf B\, =\, (3,3,1)  $ 		     & $\bf
      (3,2,1)_{(\phantom{-}\frac{1}{3},\phantom{-}\frac{1}{2})} $&
      $\bf (3,1,1)_{(-\frac{2}{3},-1)} $ & \Large$\times$  \\ \hline 
      $\bf C\,=\, (1,\overline 3, \overline 3) $ &\hspace{-.8cm}
      \begin{minipage}{11ex}\vspace{0.5ex}$\bf
      (1,2,1)_{(-1,\phantom{-}\frac{1}{2})} $ \\ $\bf
      (1,1,2)_{(\phantom{-}1,\phantom{-}\frac{1}{2})}  $
      \end{minipage}   & $\bf (1,2,2)_{(\phantom{-}0,-1)} $ &  $\bf
      (1,1,1)_{(\phantom{-}0,2)} $  \\ \hline 
    \end{tabular}
  \end{center}
  \caption{The decomposition of a $\mathbf{27}$ of $E_6$ for the
    rank 6 subgroups $SO(10)\times U(1)_\chi$, $SU(3)_C \times SU(3)_L
    \times SU(3)_R $ and their intersection $G_{LR}\times U(1)_\chi =
    SU(3)_C\times SU(2)_L\times SU(2)_R \times U(1)_{B-L}\times
    U(1)_\chi$ which is shown inside the table. The subscript denotes
    $(2\sqrt{6}/3\, \sub{Q}{B-L},\, \sqrt{6}Q_\chi )$, where the
    $U(1)$ generators are $SU(N)$ normalized.\label{tab:decomp}} 
\end{table}

Since $\mathcal{N}>1$ SUSY
does not allow chiral matter, we have reduced the overall SUSY content
to $\mathcal{N}=1$. This is achieved automatically by the orbifold
projection together with the choice of the embedding of $\mathbb Z_6$
into  the R symmetry generated by $I_R^3$, analogous to orbifold twists 
in 10D $\mathbb Z_6$ orbifolds (cf. appendix
\ref{sect:susy}). Similarly to gauge theory,  the local amount of
supersymmetry can generally be larger. In the $\mathbb Z_6$ orbifold
with 6D $\mathcal{N}=1$ SUSY in the bulk, the amount of supersymmetry
is reduced to 4D $\mathcal N=1$ on all fixed points. This is not
always the case in $T^6/\mathbb Z_6$ compactifications, for example in
the $\mathbb Z_6$-II orbifold. We assume here that possible underlying
10D shifts reduce the amount of supersymmetry to 4D $\mathcal N=1$ on
all 6D fixed points. 
Unless further constraints are taken into account, there is no {\em a
priori} rule how the three matter generations should be implemented
using the abovementioned field content. 
If they are spread over the various fixed
points, the superpotential must arise from effective nonlocal
interactions respecting the parities of the local GUT
representations. This can include scenarios where a vectorlike pair of
chiral/antichiral multiplets and another chiral multiplet combine to a
massive multiplet and a chiral multiplet in the light spectrum.
There are some plausible choices: it is for example
most appealing to implement the third generation on the fixed point with 
trinification symmetry where all fields have to be $H$-even. The localized field
content there is thus $\{{\bf A}_3,{\bf B}_3,{\bf C}_3\}$. This setup
accommodates the large top mass and has the additional advantage that
the exotics $D^c_3, D_3$ can be true leptoquarks without diquark
interactions since we can forbid the ${\bf A}^3_3$ and ${\bf B}^3_3$
terms in the superpotential by some parity. Furthermore, there is for now
only one singlet Higgs $S$ and one generation of MSSM Higgs doublets
$H_u, H_d$. The light generations must be localized on the other two
types of fixed points since here we want to implement an $H$ parity to
prevent mixing of the un-Higgs doublets (and singlets) and exotics
with standard Higgs fields and matter which would potentially produce
fatal contributions to FCNCs. If a light generation is localized on
the chiral $G_{LR}\times U(1)_\chi$ fixed point, its color charged exotics
can be leptoquarks as well, if it resides on the vectorlike $SO(10)\times
U(1)_\chi$ fixed point, they are $H$-odd and lepto/diquark interactions
are forbidden. 

We now want to use our freedom to place representations of the local gauge
groups on the fixed points as boundary localized matter, such that the
minimally required particle content of the (E)NMSSM is included, with
the simultaneous aim of obtaining unification below the Planck
scale. We have seen that the group $G_{LR}\times U(1)_\chi$ does not
unify in 4D with complete $E_6$ multiplets. One therefore might try to
work with incomplete $E_6$ multiplets. The minimal field content on
the $SO(10)\times U(1)_\chi$ invariant fixed points, ${\bf 16}_1 +
{\bf 16}_2$, could provide the matter content of the two light
generations without further exotics. Unfortunately, the standard
$U(1)_\chi$ charge assignment to matter as inherited from the $\bf
27$, $Q_\chi^{\bf 16}=1/2$, gives us 4D triangle anomalies which are only
cancelled by the exotics with $Q_\chi^{\bf 10}=-1$ and $Q_\chi^{\bf
  1}=2$ in the full multiplet. One might consider assigning a special
charge $Q_\chi=-1/2$ to the first generation matter to make the
incomplete matter vector-like under $U(1)_\chi$. This cancels the
$SO(10)\times U(1)_\chi$ anomalies, but does not allow a
superpotential at tree-level, and requires breaking $U(1)_\chi$ at a
high scale to generate it. We therefore use the full three generation
particle content and achieve unification below the Planck scale by
different means. 

\begin{figure} \begin{center}
     \includegraphics[width=0.48\linewidth]{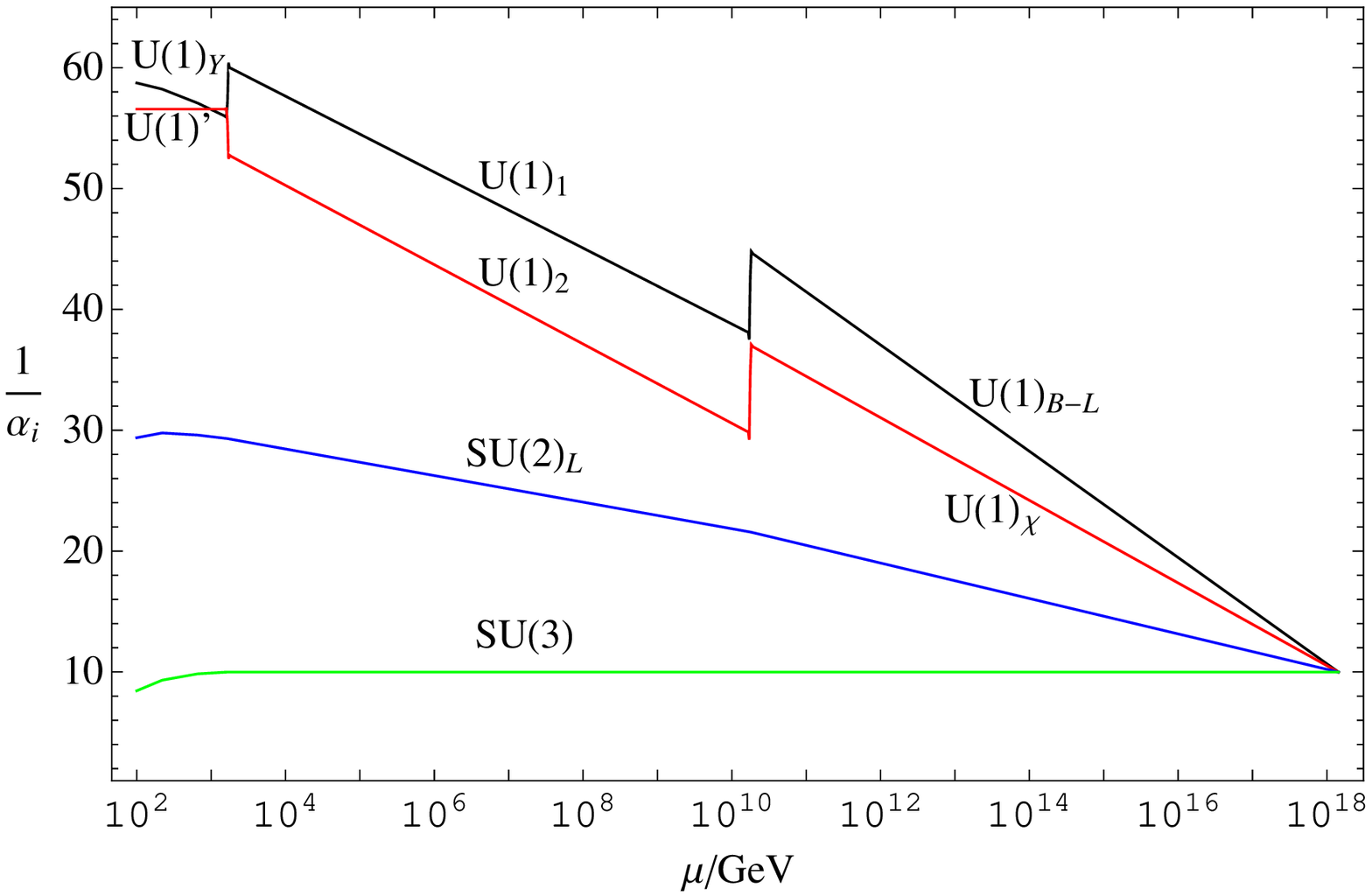}
     \includegraphics[width=0.48\linewidth]{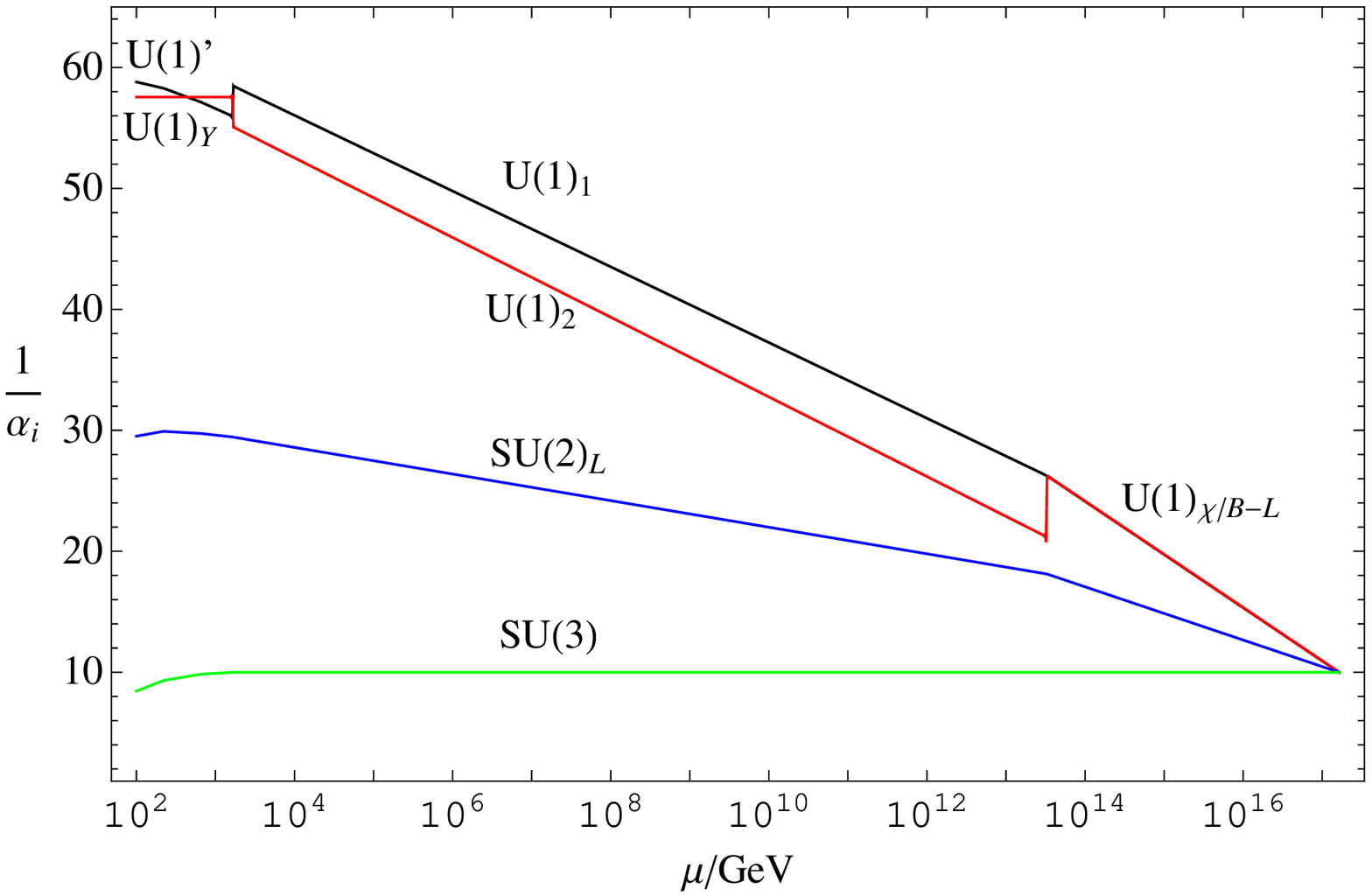}\\
     \includegraphics[width=0.48\linewidth]{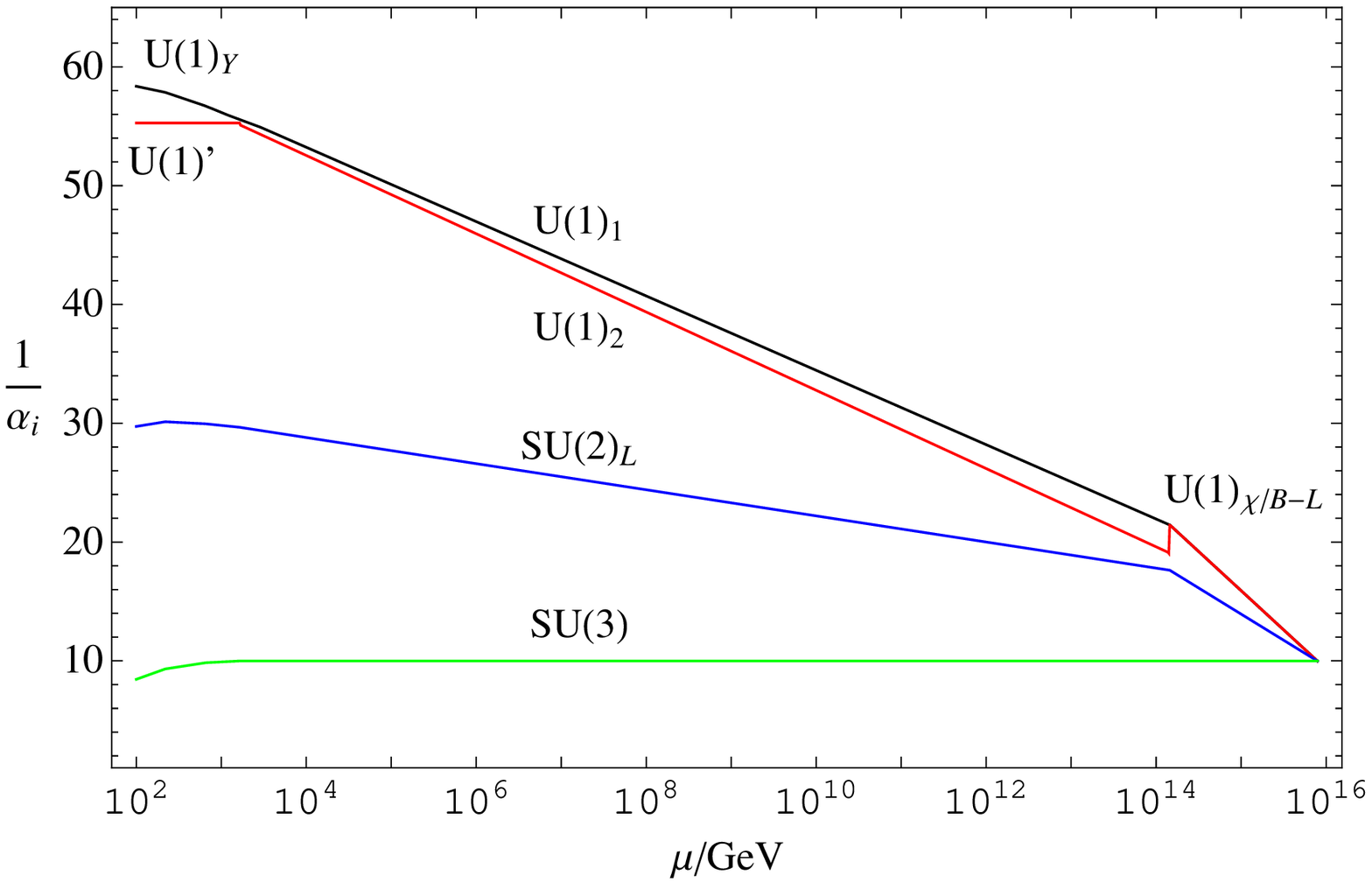}
     \includegraphics[width=0.48\linewidth]{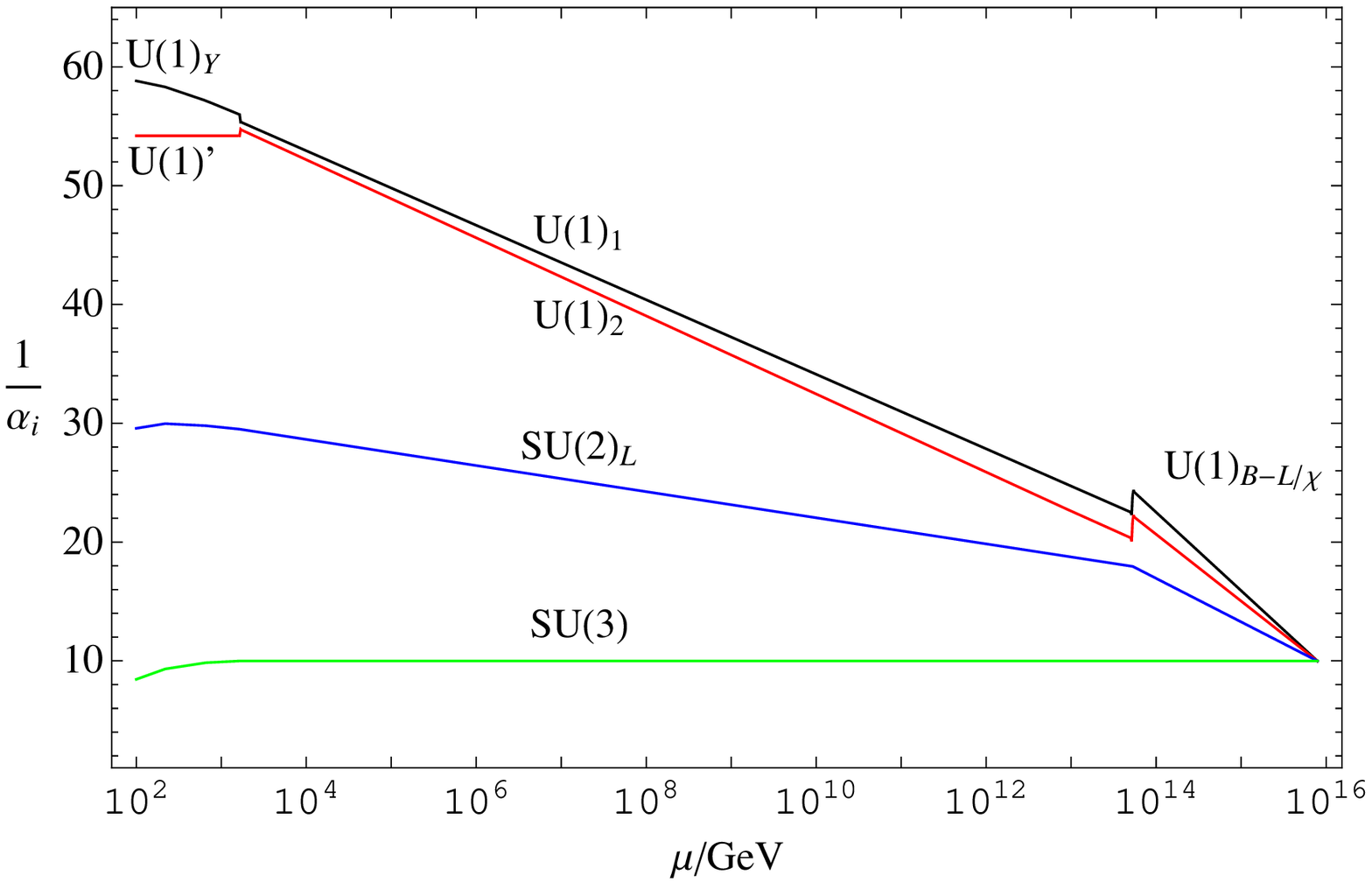}
\end{center}
\caption{Four unification scenarios at 1-loop with matter in complete $E_6$
multiplets and intermediate breaking with combinations of incomplete vector-like
Higgs representations $i)$ and $ii)$ as allowed by the orbifold
compactification. The cases are $i), ii), 3 \times ii)$ and   $2\times
ii)+1\times i)$. Below the intermediate scale, the mixing of the Abelian gauge
groups is taken into account. \label{fig:unifyincomplete}} \end{figure}

To calculate the unification scenario, we need as one further
ingredient the breaking mechanism of the intermediate symmetry, 
\begin{equation}
G_{LR}\times U(1)_\chi \longrightarrow SU(3)_C \times SU(2)_L \times
U(1)_Y\times U(1)'. 
\end{equation}
Giving some Higgs fields $\sub{H}{int},\,\sub{\bar{H}}{int}$ vevs in
$\tilde{\nu}^c$ and $\tilde{\overline{\nu}}^c$  directions
respectively at $\sub{\Lambda}{int}$ leaves arbitrary linear
combinations of 
\begin{align}
  g_Y Y&=\frac{g_R g_{B-L} \left(\sqrt{3}  T^3_R+\sqrt{2}
    Q_{B-L}\right)}{\sqrt{3 g_{B-L}^2+2 g_R^2}} \hspace{4cm}
  \mbox{and}\\ 
  g^\prime Q^\prime &=
  \frac{g_{\chi } \left(g_R^2 \left(2 \sqrt{3} Q_{\chi }-\sqrt{2}
    T^3_R\right)+\sqrt{3} g_{B-L}^2 \left(Q_{B-L}+3 
    Q_{\chi }\right)\right)}{\sqrt{\left(3 g_{B-L}^2+2 g_R^2\right)
  \left(9 g_{B-L}^2+6 g_R^2+g_{\chi }^2\right)}} 
\end{align}
invariant, yielding the matching conditions for the corresponding
coupling constants via the GUT normalization of the charges 
\begin{equation}
g^Q=\sqrt{2 \tr_a[(g^Q Q_a)^2]}.
\end{equation}
Like in models with intermediate Pati-Salam symmetry, this can be accomplished
by using a $\bf 27+ \overline{27}$ which does not lead to unification as
demonstrated in Fig.~\ref{fig:LRwo27}. In our case, we can consider to
use parts of $\bf 27 + \overline{27}$, namely $\bf 16 + \overline{16}$
on the $SO(10)\times U(1)$ fixed points, an $\bf
(1,\overline{3},\overline{3})+(1,3,3)$ on trinification fixed points, 
or a representation of $G_{LR}\times U(1)_\chi$ as given in Table \ref{tab:decomp} (alternatively, one can consider
putting an $\mathbf{(1,1,3) + (1,1, \overline 3)}$ on a trinification fixed point which is not
contained in a small $E_6$ multiplet and has different quantum
numbers). For these breaking scenarios to be viable, potentials need to be
found which produce these vevs in a $D$-flat
direction. For further analysis, from here on we assume for our
unification scenario that there is no additional matter content below
$\sub{\Lambda}{int}$. Two choices for the additional matter doing the
job of the intermediate symmetry breaking are
\begin{eqnarray}
  \label{intHiggs}
  i) && L, l^c, \langle \nu^c\rangle + c.c.  \sim  {\bf (1,\overline
    3,\overline 3)} \cap  {\bf 16} + c.c. \\ 
  ii) && L, l^c,\langle \nu^c\rangle , H_u, H_d,S +  c.c.  \sim {\bf
    (1,\overline 3,\overline 3)} + c.c. 
\end{eqnarray}
These representations affect the running between $\Lambda_{int}$ and
$\Lambda_{E6}$, and would correspond to putting the intermediate Higgs
on a $G_{LR}\times U(1)_\chi$ invariant fixed point or the more
symmetric trinification localized case, respectively. Since we are
free to split heavy vector-like $\bf27 + \overline{27}$ at
$\Lambda_{E6}$, these choices do not necessarily correspond to really
incomplete multiplets. Some examples for the running with such a
breaking matter content are shown in
Fig.~\ref{fig:unifyincomplete}. These scenarios unify  
below (though in one case close to) the Planck scale, and since in $\delta$ extra dimensions we can roughly
identify the extra dimensional volume with $V_{\delta}^{}\sim\Lambda_{E6}^{-\delta}$, they also unify
below the fundamental gravity scale $M_*$ because of
$M_*/\Lambda_{E6}\sim (\Lambda_{Pl}^{} / \Lambda_{E6})^{{2}/{(2+\delta)}}$. 
The unified coupling at the $E_6$ scale is
$\alpha(\Lambda_{E6})\approx \alpha_s(m_Z)\approx 0.1$. Those scenarios with
larger intermediate particle content tend to unify at a higher $E_6$ scale and
to have a lower intermediate breaking scale $\Lambda_{\mbox{\tiny{int}}}$. 
The exact $E_6$ unification which we assume here for simplicity is in general modified by threshold 
corrections from the local GUT compactification, which results in a modified 
intermediate breaking scale. The analysis of the impact of these corrections to unification and
low energy observables is beyond the scope of this work, but must be taken into 
account in order to precisely determine the breaking scenario from measurements of couplings
at the TeV scale.
Below $\Lambda_{\mbox{\tiny{int}}}$, the right-handed neutrino is integrated
out, leaving a matter content in incomplete $\mathbf{27}$, which results in a
mixing of the two Abelian gauge groups $U(1)_Y$ and $U(1)^\prime$ at the 1-loop
level.  However, there is a basis $\{U(1)_1, U(1)_2\}$ in which the two
couplings run independently. At the scale, where the NMSSM-like singlet $S$
acquires a vev, $U(1)_Y$ is projected out as the unbroken linear combination of
$U(1)_1$ and $U(1)_2$.
The broken generator corresponds to $U(1)^\prime$, resulting in a
heavy $Z^\prime$ boson with mass of around a TeV.  
Its coupling to matter is roughly of the strength of the SM
hypercharge at the $Z$ pole. A list containing the numerical values of the
$U(1)^\prime$ couplings and charges, corresponding to the four
scenarios presented in Fig.~\ref{fig:unifyincomplete}, can be found
in 
Table \ref{chargelist}. It lists the charges and coupling strengths of the
matter coupling to the $Z^\prime$ boson. At this stage, threshold corrections from orbifold effects are not yet taken into account. 
As $Q^\prime$ depends on the
ratio of the gauge couplings at $\sub{\Lambda}{int}$, the charges are
generally non-rational. 
\begin{table}[h]
  \begin{center}
    \begin{tabular}{|c|cccc|}
      \hline
      $\sub{H}{int},\sub{\bar{H}}{int} $&$ i)$ &$ii)$&$3ii)$ &$i)+2ii)$\\
      \hline
      $\sub{\Lambda}{int}/$GeV &$ 1.6\times10^{10}$ &$ 3.0\times10^{13}$ &
      $ 1.3\times10^{14}$ &$ 4.9\times10^{13}$\\
      $\sub{\Lambda}{GUT}/$GeV &$ 1.3\times10^{18}$ &$ 1.5\times10^{17}$ &
      $ 7.2\times10^{15}$ &$ 7.2\times10^{15}$\\
      $g^\prime|_{M_{Z^\prime}}$& 0.471& 0.467 &0.476& 0.482\\
      \hline 
      $Q^\prime_X $&&&&\\ 
      \hline
      $Q $  & 0.224 & 0.231 & 0.234 & 0.232 \\
      $u^c $  & 0.283 & 0.261 & 0.250 & 0.257 \\
      $d^c $  & 0.055 & 0.067 & 0.073 & 0.069 \\
      $D $  & -0.449 & -0.462 & -0.468 & -0.464 \\
      $D^c $  & -0.339 & -0.328 & -0.322 & -0.326 \\
      $L $  & 0.114 & 0.097 & 0.089 & 0.094 \\
      $e^c $  & 0.165 & 0.201 & 0.218 & 0.208 \\
      $H^u $  & -0.508 & -0.492 & -0.484 & -0.489 \\
      $H^d $  & -0.279 & -0.298 & -0.307 & -0.301 \\
      $S $  & 0.787 & 0.790 & 0.790 & 0.790\\
      \hline
    \end{tabular}
    \caption{$U(1)^\prime$ charges and couplings corresponding to the four
      scenarios from Fig.~\ref{picorbiz2}. $i)$ and $ii)$ label the
      matter content as in eq. (\ref{intHiggs}) which breaks the
      intermediate gauge group. \label{chargelist}}
  \end{center}
\end{table}

\subsection{Low-energy scenario and phenomenological aspects}
Decomposing the superpotential from (\ref{W-trini}) under SM $\times U(1)^\prime$, one obtains:
  \begin{align}
    W\,&= \,Y^u\,u^c\,Q\,H^u\,
        + \,Y^d\,d^c\,Q\,H^d\,+\,
          Y^e\,e^c\,L\,H^d\,\notag 
      \\&+\,
          Y^D\,D\,u^c\,e^c\,+\,
          Y^{D^c}\,D^c\,L\,Q\,+\,
          Y^{SD}\,S\,D^c\,D\,+\,
          Y^{SH}\,S\,H^u\,H^d\;.
  \end{align}
Note, that this already incorporates $R$-parity since $U(1)_{\mbox{\tiny{B-L}}}$
is contained in $E_6$. In general, for example if the corresponding matter originates from fixed points where
$E_6$ is broken to $H\subset SO(10)\times U(1)_\chi$, the trinification relations between
$Y^{SD}$, $Y^{SH}$ and $Y^e$, and between $Y^u$, $Y^d$,$Y^D$, $Y^{Dc}$ are relaxed, which allows the superpotential to respect $H$ parity.
If at least one generation of color charged exotics $D, D^c$ have leptoquark couplings (and are thus $H$-even) as it is allowed by scenarios with trinification 
or $G_{LR}\times U(1)_\chi$ intermediate breaking, it is conceivable that the $H$-odd exotics can decay into $H$-odd singlets and $H$-even leptoquarks. This might
circumvent the need of models with intermediate $PS\times U(1)_\chi$ and no leptoquarks for a small $H$ parity breaking if such a scenario can be brought to be consistent with cosmological bounds on neutral dark matter.

Electroweak symmetry breaking is triggered by the
$S$ field obtaining a vev which appears to be naturally induced via radiative
effects, as its coupling to the exotics drives the soft-breaking mass parameter
$m^2_S$ to negative values. The quartic term in $S$ originates from the $U(1)^
\prime$ $D$ term, linking the vev of the singlet to the mass of the $Z^\prime$ 
boson. 
In order to be consistent with experimental data, the singlet vev has to be
above one TeV, generically decoupling the MSSM-like Higgs sector from the 
singlet field.


\section{Conclusions and Outlook}

We have demonstrated how $\mathbb Z_n$ orbifold constructions can be
used to generate an  $E_6$ inspired Exceptional SSM with a $D$ term
induced singlet vev, circumventing many problems of comparable
standard $E_6$ GUT constructions. While models with intermediate $PS\times U(1)_\chi$
symmetry can be constructed in 5D orbifolds, the aim to have intermediate $LR$ symmetry
led us to consider 6D geometries. The resulting models potentially
provide a rich phenomenology at collider experiments: In addition to
the generic $R$-parity conserving supersymmetric signatures, a
$Z^\prime$ boson with a mass around the TeV scale as well as strongly
produced leptoquark-like exotics should guarantee a high discovery
potential at the LHC. This is in contrast to $SU(5)$ derived standard
GUT unification scenarios which offer no directly observable collider 
phenomenology beyond the MSSM. 

Apart from $R$-parity, there is $H$-parity rendering the lightest un-Higgs 
or un-Higgsino stable (LHP). If the lightest particle which is odd
under both parities is not heavy enough to decay into
lighter singly-odd particles, there may even be a third type of
dark matter. As there are multiple ways from the resulting
multicomponent relic densities to be in agreement with current bounds
from WMAP and astrophysical observations, they can add interesting
aspects to the interpretation of recent direct and indirect WIMP searches.

On the theoretical side, it would be interesting to explore whether such or
similar scenarios can be embedded in complete heterotic string models. Those
impose strict additional constraints and thus can reduce the level of
arbitrariness in the choice of Wilson lines, representations and the
massless spectrum which is inherent to the effective field theory
approach. In this context it should also be possible to address the
generation of the effective superpotential needed if matter is
localized on different fixed points. This is postulated in the field
theory construction, and we have only sketched this in this paper. It is also
noteworthy that the 1-loop QCD beta function vanishes due
to the color charged light exotics, and the unified gauge coupling in $SU(N)$
normalization, $\alpha(\Lambda_{E6})\approx 0.1$, is therefore considerably
larger than in standard MSSM unification.


\section*{Acknowledgements}

We would like to thank Patrick Vaudrevange and Pavel Fileviez Perez
for enlightening discussions. This work has been partially funded by
the DFG under grant No. RE 2850/1-1 and the Graduate School GRK 1102
``Physics at Hadron Colliders'' as well as by the Juniorprofessor
program of the Ministry of Science and Culture (MWK) of the German
state Baden-W\"urttemberg. 


\appendix
\section{LR symmetric $E_6$ breaking on a Circle and a Torus}

\subsection{Shift embeddings for $\mathbb Z_2$, $\mathbb Z_3$, $\mathbb Z_4$ and $\mathbb Z_6$} 
\label{sec:embeddings}
Since we are interested in LR symmetric intermediate groups of rank 6, we
restrict the orbifold action in group space to those generated by the Cartan
generators.  Let $\theta$ be a generator of an orbifold symmetry, e.g.
rotations, translations or parities. We can now associate with it an action in
group space which we parameterize 
with a shift vector $V$ acting on the roots of $E_6$. The roots thus transform
as 
\begin{equation}
  G(\theta) E_\alpha = \exp\left[2 \pi i  V \cdot H \right] E_\alpha  =
  \exp\left[2 \pi i  V \cdot \alpha \right]E_\alpha 
\end{equation}
where we demand that $V$ is chosen to be compatible with the multiplication law
of the orbifold space group, i.e.  \begin{equation}G(\theta)^n=\exp\left[2 n
\pi i  V \cdot H \right]=1  \end{equation} for $\mathbb Z_n$. This is
equivalent to
\begin{equation}
  \forall \alpha: V\cdot \alpha  \in \mathbb Z/n 
\end{equation}
and $nV$ should thus be element of the co-root lattice. The unbroken
subgroup which survives under such an orbifold action is given by the
set of generators (the Cartan subalgebra and roots) which are left
invariant under the action of $\theta$. 

To narrow down the number of candidates, we demand that the minimal viable LR
symmetric subgroup of rank 6, $G_{LR}\times U(1)_\chi $,
in the embedding given in Table \ref{tab78} is contained in the
invariant part $H \subset E_6$. The resulting groups and representative shift
vectors are shown in Table \ref{tabzshifts}.  For $\theta^2=1$, we find the two
largest subgroups. 
If we extend our reach to transformations which obey $\theta^3=1$, we find the
trinification group. The case $\theta^4=1$ includes the $\mathbb Z_2$
candidates, and two new groups containing $SU(3)_{L/R}$.  Combining two
orbifold parities, we find the common subgroups given in Table
\ref{tabzzhifts}. Since the generators which survive on any fixed point or line
are even under $\theta$, we have to define boundary localized matter to be even
under the orbifold twist as well. Likewise, of any matter
representation with weights $\mu_i$, which transforms as (e.g. bulk matter)

\begin{equation}G(\theta)| \mu_i \rangle=  e^{2 k \pi i/n}\exp\left[2
    \pi i  V \cdot \mu_i \right]|\mu_i\rangle,\quad k=1,\dots, n \;\; ,
\end{equation} 
only the even parts remain on the fixed
point.  

\begin{table}
  \begin{center}
%
%
    \begin{tabular}{|l|l|l|}
      \hline
      $\mathbb Z_2$&Subgroup $H$& Shift $2\overline{V}$ \\
      \hline
      &$SO(10)\times U(1)_\chi$ & $(1,1 ,0 ,1 ,1 ,0  )$\\
      \hline
      &$SU(6)\times SU(2)_R$ & $(0 ,0 ,1 ,0 ,0 ,0 )$\\
      \hline
      &$SU(6)\times SU(2)_L$ & $(1 ,1 ,1 ,1 ,1 ,0 )$\\
      \hline \hline
      $\mathbb Z_3$&   Subgroup $H$& Shift $3\overline{V}$ \\
      \hline
      &$SU(3)_C\times SU(3)_L\times SU(3)_R$ & $(0,0,1,2,0,0 )$\\
      \hline \hline
      $\mathbb Z_4$& Subgroup $H$& Shift $4\overline{V}$ \\ \hline
      &$SU(5)\times U(1)\times SU(2)_L $ & $(3,1,3,1,1,0)$ \\ \hline
      &$SU(5)\times U(1) \times SU(2)_R $ & $(2,2,1,0,2,0 )$ \\ \hline
      &$SU(4)_C\times SU(2)_L \times SU(2)_R \times U(1)_\chi$ & $(3,1,2,3,1,0)$ \\ \hline
      &$SU(3)_C \times SU(3)_L\times SU(2)_R \times U(1)$ & $(0,0,1,2,0,0)$\\
      \hline
      &$SU(3)_C \times SU(3)_R\times SU(2)_L \times U(1)$ & $(3,1,1,1,1,0)$\\
      \hline \hline
      $\mathbb Z_6$& Subgroup $H$& Shift $6\overline{V}$ \\ \hline
      &$SU(3)_C \times SU(3)_L\times SU(2)_R \times U(1)$ & $(4,2,1,0,2,0)$\\
      \hline
      &$SU(3)_C \times SU(3)_R\times SU(2)_L \times U(1)$ & $(5,1,5,3,1,0)$\\ \hline
      &$SU(3)_C \times SU(2)_L\times SU(2)_R \times U(1)_{B-L}\times U(1)_\chi$ & $(1,5,4,3,5,0)$\\ \hline
    \end{tabular}
\end{center}
  \caption{The subgroups $E_6 \supset H \supset G_{LR}\times U(1)_\chi$
    invariant under shifts which are compatible with $\mathbb Z_2$,
    $\mathbb Z_3$, $\mathbb Z_4$ amd $\mathbb Z_6$. Representative shift vectors for
    each case are given in the dual basis such that $V\cdot
    \alpha=\overline{V}\cdot \Delta(\alpha)$. \label{tabzshifts}} 
\end{table}
\begin{table}
  \begin{center}
    \begin{tabular}{|l|l|}
      \hline  
      $\mathbb Z_2 \times \mathbb Z_2  $& $SU(4)_C\times SU(2)_L \times SU(2)_R \times U(1)_\chi$ \\ \hline
      $\mathbb Z_2 \times \mathbb Z_3  $& $SU(3)_C\times SU(2)_L \times SU(2)_R \times U(1)_{B-L} \times U(1)_\chi $ \\ 
      & $SU(3)_C \times SU(3)_L \times SU(2)_R \times U(1) $ \\ 
      & $SU(3)_C \times SU(3)_R \times SU(2)_L \times U(1) $ \\ \hline
      $\mathbb Z_2 \times \mathbb Z_4  $ &  $SU(4)_C\times SU(2)_L \times SU(2)_R \times U(1)_\chi $  \\ 
      &  $SU(3)_C\times SU(2)_L \times SU(2)_R \times U(1)_{B-L}\times U(1)_\chi $  \\ \hline 
      $\mathbb Z_3 \times \mathbb Z_4  $& $SU(3)_C \times SU(3)_L \times SU(2)_R \times U(1) $  \\ 
      & $SU(3)_C \times SU(3)_R \times SU(2)_L \times U(1) $ \\ 
      & $SU(3)_C \times SU(2)_L \times SU(2)_R \times U(1)_{B-L}\times U(1)_\chi $ \\ \hline
   $\mathbb Z_4 \times \mathbb Z_4$ &  $SU(4)_C\times SU(2)_L \times SU(2)_R \times U(1)_\chi $\\
& $SU(3)_C \times SU(2)_L \times SU(2)_R \times U(1)_{B-L}\times U(1)_\chi $
      \\ \hline
    \end{tabular}
  \end{center}
  \caption{The nontrivial ($H_i \nsubseteq H_j$) common invariant
    subgroups $H_i \cap H_j$ under combinations of two
    shifts.\label{tabzzhifts}} 
\end{table}

\subsection{Orbifold Geometries in $D=5$}\label{sec:5dgeometries}
In $D=5$ there are essentially two orbifolds, $S^1 / \mathbb Z_2$ and
$S^1 / \mathbb (\mathbb Z_2 \times \mathbb Z'_2)$, which have as
fundamental domains an interval $[0,\pi R]$ with boundary conditions
imposed on both ends (corresponding to the fixed points). The former can be seen as a special case of the latter 
without discrete Wilson line. An exemplary construction is shown in Figure \ref{fig:s1orbifold}. One can start with the compactification of the real line $\mathbb R$
to $S^1$ by modding out a translation $t$. An order 2 reflection about the origin $r$ induces a second reflection $r'=t r$. The fundamental domain is an interval with
the fixed points of $r$ and $r'$ as boundaries. One can now assign different $\mathbb Z_2$ gauge shifts to the two reflections $r$ and $r'$ to break $E_6$ down to subgroups
on the two fixed points, resulting in a discrete Wilson line.

\subsection{Orbifold Geometries in $D=6$}\label{sec:6dgeometries}
We now want to consider the possibilites how $E_6$ can be broken to 
suitable rank 6 subgroups containing $G_{LR}\times U(1)_\chi$ using two extra dimensions.
In general, we can consider $\mathbb R^2 / \Gamma$ where
$\Gamma$ is one of the 17 plane crystallographic groups. To obtain a 
gauge symmetry close to $G_{PS}\times U(1)_\chi$ or $G_{LR}\times U(1)_\chi$ and hence less constrained
superpotentials, we search for 6D geometries with fixed points in
which two generators, either rotations or
translations, can carry nontrivial gauge embeddings. In this paper, we consider only
rotational embeddings, and leave the possibilities of geometries involving mirror 
planes such as the orbifolds $\bf 3\!*\!3$ and $\bf *333$ for future investigation\footnote{For the crystallographic groups we use the
  orbifold notation, cf. \cite{conway}.}.
We discuss some examples which can be useful in LR symmetric $E_6$ breaking.

\subsubsection{Maximal Subgroups from a $\mathbb Z_2$ Orbifold}

\begin{figure}
  \begin{center}
    \raisebox{-3ex}{\includegraphics[width=5cm]{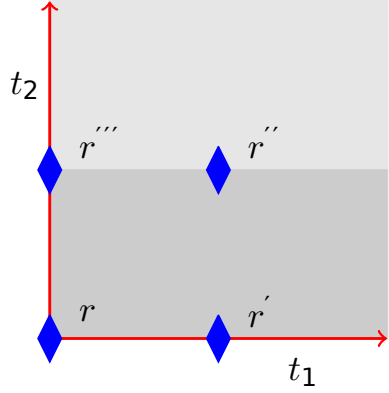}}
  \end{center}
  \caption{The $\mathbb R^2 / {\bf 2222}$ orbifold which is one of the
    $T^2/\mathbb Z_2$ orbifolds. The blue diamonds indicate 
    the fixed points under the four $180^\circ$ rotations, while the red
    arrows are the translations which span the torus. The 
    light shaded area indicates the torus, the dark shaded area
    corresponds to the fundamental domain of the
    orbifold.\label{picorbiz2}}  
\end{figure}
The $T^2/\mathbb Z_2$ geometry $\mathbb R^2 / {\bf
  2222}$ as
shown in Fig.~\ref{picorbiz2} is generated by modding out a torus
(not necessarily a $90^\circ$ one as shown here) with a $180^\circ$
(order 2) rotation $r$ about the origin, to which we can assign a shift
embedding 
\begin{equation}
  G(r)E_\alpha \in \{1,-1 \}E_\alpha \; .
\end{equation}
Together with the two translations $t_1$ and $t_2$ denoted by the red
arrows, this induces four rotations of order 2 shown as blue
diamonds. They are generated by  
\begin{eqnarray}
  r'= t_1 r, \quad
  r'' = t_2 t_1 r, \quad
  r''' = t_2 r
\end{eqnarray}
The orbifold offers the possibility to assign two $\mathbb
Z_2$-discrete Wilson lines to $t_1$ and $t_2$. The resulting gauge
shifts 
are 
\begin{eqnarray}
  G(r')E_\alpha  &=& G(t_1) G(r)E_\alpha\in \{1,-1\}E_\alpha\\
  G(r'')E_\alpha &=& G(t_2)G(t_1)G(r)E_\alpha\in \{1,-1\}E_\alpha \\
  G(r''')E_\alpha &=& G(t_2) G(r)E_\alpha\in \{1,-1\}E_\alpha 
\end{eqnarray}
As was summarized in Tables~\ref{tabzshifts} and \ref{tabzzhifts}, we
can use these assignments to break $E_6$ to either $SO(10)\times U(1)_\chi$,
$SU(6)\times SU(2)_L$, $SU(6)\times SU(2)_R$ or the intersection of
either pair, $SU(4)\times SU(2)_L \times SU(2)_R \times
U(1)_\chi$. The fixed points themselves always exhibit the local gauge
invariance $E_6$, $SO(10)\times U(1)_\chi$ or $SU(6)\times SU(2)$. 

\subsubsection{Trinification on a $\mathbb Z_3$ Orbifold}

\begin{figure}
  \begin{center}
    \includegraphics[width=7cm]{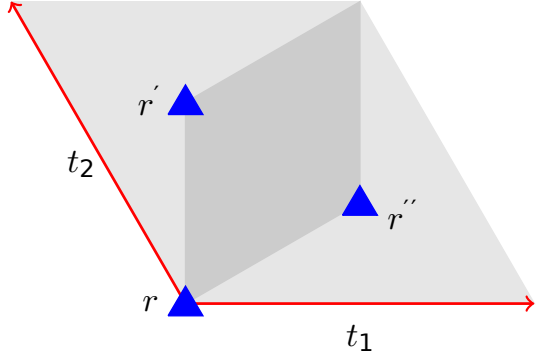}\vspace{-1ex}
  \end{center}
  \caption{The $\mathbb R^2 / {\bf 333}$ orbifold which is one of the
    $T^2/\mathbb Z_3$ orbifolds. The blue triangles indicate 
    the fixed points under the three $120^\circ$ rotations, while the
    red arrows are the translations which span the torus. The 
    light shaded area indicates the torus, the dark shaded area
    corresponds to the fundamental domain of the
    orbifold.\label{picorbip3}}  
\end{figure} 

We can obtain the trinification group $SU(3)_C \times SU(3)_L
\times SU(3)_R$ by modding out the torus, $T^2/\mathbb Z_3$,
corresponding to the geometry $\mathbb R^2 / \mathbf{333}$ as shown in
Fig.~\ref{picorbip3}. The fundamental space is a 3-pillow. It is
generated by a $120^\circ$ rotation $r$ about the origin, which we can
give a $\mathbb Z_3$ shift embedding,  
\begin{equation}
  G(r)E_\alpha \in \{1, e^{2\pi i /3}, e^{4 \pi i /3}\}E_\alpha \; .
\end{equation}
In addition, we can assign discrete Wilson lines to the translations
which span the torus, which induce different gauge shifts for the two
rotations $r'=t_1 t_2 r$ and $r''=t_1 r$, 
\begin{eqnarray}
  G(t_1)E_\alpha = G(t_2)E_\alpha \equiv G(t)E_\alpha\in \{1, e^{2\pi
  i /3}, e^{4 \pi i /3}\}E_\alpha\\ 
  G(r') = G(t)^2G(r),\quad G(r'') = G(t) G(r) = G(t)^{-1}G(r')
\end{eqnarray}
This gives us the possibility to have reduced gauge symmetry $SU(3)^3$
on none, two or all three corners of the pillow. We can choose for
example 
\begin{eqnarray}
  G(r)E_\alpha&=&\exp\Big[2 \pi i\,
    (0,0,\frac{1}{3},-\frac{1}{3},0,0)\cdot
    \Delta(\alpha)\Big]E_\alpha,\quad G(t)=1 
\end{eqnarray}
as given in Table \ref{tabzshifts}.
The resulting invariance $SU(3)^3$ contains the $G_{LR}\times U(1)_\chi$
subgroup which can for example be obtained if an additional vev in the
direction $t_{B-L}$ is present. Since all fixed points are invariant
under $SU(3)^3$ only, it is interesting to consider what is the most
general invariant superpotential for each. Inspired by the
anomaly-free $E_6$ particle content, we start with the decomposition
of the $\mathbf{27}$ under the trinification group, which is 
\begin{eqnarray} 
  \bf 27 &\rightarrow & \bf \underbrace{\bf(\overline 3, 1,3)}_{A} +
  \underbrace{\bf(3,3,1)}_{B} + \underbrace{\bf(1,\overline
  3,\overline 3)}_{C} 
\end{eqnarray}
where $\mathbf{A}$ contains right-handed quarks $q_R^c$ and an exotic
$D^c$, $\mathbf{B}$ contains left-handed quarks $q_L$ and an exotic
$D$, while $\mathbf{C}$ contains all Higgs $H_u$, $H_d$ fields including
the SM singlet $S$, and all leptons. The $E_6$-invariant
renormalizable trilinear superpotential decomposes into four
independent $SU(3)^3$ invariants, 
\begin{eqnarray} 
  \bf 27^3 &\rightarrow& c_1 {\bf A}^3 + c_2 {\bf B}^3 +c_3 {\bf C}^3
  + c_4 {\bf ABC} \; ,
  \label{W-trini}
\end{eqnarray}
which is already the most general renormalizable superpotential one can write
down for this field content.  The first term ${\bf A}^3$ contains only
diquark-like couplings to $D^c$, while ${\bf B}^3$ only contains diquark-like
couplings to $D$. The remaining two invariants give us a complete MSSM-like
superpotential including an effective $\mu$ Term $S H_u H_d$, leptoquark
couplings and effective leptoquark masses $S D^c D$. If we demand $c_1=c_2=0$
by some symmetry, $D$ and $D^c$ thus become true leptoquarks. Here, any
symmetry will do under which $\mathbf{A}$ and $\mathbf{B}$ are odd
while $\mathbf{C}$ is uncharged such as baryon number or parity. Even
with intermediate breaking to $G_{SM} \times U(1)$ for example from an adjoint vev ${\bf \langle 78\rangle} \sim Q_{B-L} $, none of the
$SU(3)^3$ subgroups unify below the Planck scale assuming the full
$E_6$ particle content (with right-handed neutrinos massive at the
intermediate scale). Therefore, one has to invoke additional physics
to achieve unification, either incomplete $E_6$ multiplets or 
extra-dimensional effects \cite{Hebecker:2002vm,Hebecker:2004xx}.


\subsubsection{Pati-Salam Symmetry on a $\mathbb Z_4$ Orbifold\label{sect:geometry442}}
\begin{figure}
\begin{center}
  \includegraphics[width=5cm]{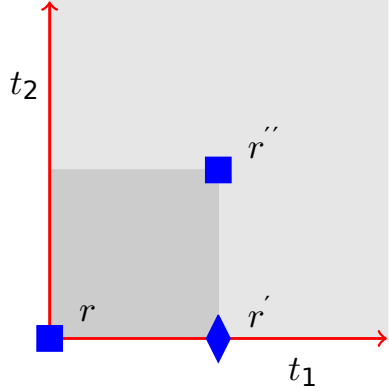}
\vspace{-3ex}
\end{center}
\caption{The $\mathbb R^2 / {\bf 442}$ orbifold which is a
  $T^2/\mathbb Z_4$ orbifold. The blue rectangles and and diamond
  indicate the fixed points under the $90^\circ$ and $180^\circ$ 
  rotations respectively, while the red arrows are the
  translations which span the torus. The light shaded area indicates
  the torus, the dark shaded area corresponds to the fundamental
  domain of the orbifold.\label{picorbip4}} 
\end{figure} 

The simplest 6D orbifold symmetry without reflections which can accomodate $PS\times U(1)_\chi$ fixed points is the $T^2/\mathbb Z_4$ orbifold $\mathbb R^2/{\bf 442}$. It is shown in Fig.~\ref{picorbip4}. The order 4 rotation $r$ together with the translations $t_1$ and $t_2$ induce an order 4 and an order 2 rotation
\begin{eqnarray}\label{eqn:z4rotations}
r'' = t_1 r, \quad r' = t_1 r^2
\end{eqnarray}
where $r t_1 = t_2 r $. There  is one discrete Wilson line $G(t)=G(t_1)=G(t_2)$ of order two. To obtain a model with 4D gauge invariance $PS\times U(1)_{\chi}$
with fixed points that allow H parity and the suppression of lepto-diquark couplings, we can for example choose  
\begin{eqnarray}
  G(r)E_\alpha&=&\exp\Big[2 \pi i\,
    (\frac{3}{4},\frac{1}{4},\frac{1}{2},\frac{3}{4},\frac{1}{4},0)\cdot
    \Delta(\alpha)\Big]E_\alpha\nonumber \\ 
    G(t)E_\alpha&=&\exp\Big[2 \pi i\,
    (\frac{1}{2},\frac{1}{2},0,\frac{1}{2},\frac{1}{2},0)\cdot
    \Delta(\alpha)\Big]E_\alpha 
\end{eqnarray}
as given in Table \ref{tabzshifts}. Now, $E_6$ is broken down to $PS\times U(1)_\chi$ at the order 4 fixed points and remains unbroken at the order 2 fixed point. If one chooses instead $G(t)=1$, it is broken to $SO(10)\times U(1)_\chi$ at the order 2 fixed point, allowing H parity.
\subsubsection{Minimal LR Symmetry on a $\mathbb Z_6$ Orbifold\label{sect:geometry632}} 

\begin{figure}
\begin{center}
  \includegraphics[width=7cm]{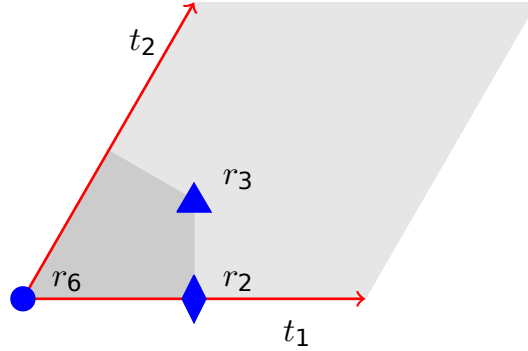}
\vspace{-3ex}
\end{center}
\caption{The $\mathbb R^2 / {\bf 632}$ orbifold which is one of the
  $T^2/\mathbb Z_6$ orbifolds. The blue circle, triangle and diamond
  indicate the fixed points under the $60^\circ$, $120^\circ$ and
  $180^\circ$ rotations respectively, while the red arrows are the
  translations which span the torus. The light shaded area indicates
  the torus, the dark shaded area corresponds to the fundamental
  domain of the orbifold.\label{picorbip6}} 
\end{figure} 

The simplest 6D orbifold geometry which can contain both nontrivial
$\mathbb Z_2$ and $\mathbb Z_3$ shifts is the $T^2/\mathbb Z_6$
orbifold $\mathbb R^2/ {\bf 632}$ or p6. It is shown in
Fig.~\ref{picorbip6}. The order 6 rotation $r_6$ together with the 
translations $t_1$ and $t_2$ induces an order 2 and an order 3
rotation 
\begin{eqnarray}
\label{eqn:z6rotations}
r_2 = t_1 r_6^3, \quad r_3 = t_1 r_6^2  \; .
\end{eqnarray}
Although we are not allowed to assign discrete Wilson lines since from
equation~(\ref{eqn:z6rotations}) follows that $G(t_1)^2 =
G(t_1)^3=\fatone$, we can choose gauge shifts for the $\mathbb Z_2$
and $\mathbb Z_3$ rotation separately.  
The possible shifts satisfy
\begin{eqnarray}
G(r_2)E_\alpha&=&G(r_6)^3 E_\alpha \in \{1, -1\}E_\alpha\\
G(r_3)E_\alpha &=& G(r_6)^2 E_\alpha\in \{1, e^{2\pi i /3}, e^{4 \pi i /3}\}E_\alpha
\; ,
\end{eqnarray}
and vice versa
\begin{eqnarray}
G(r_6)E_\alpha&=&G(r_2) G(r_3)^{-1}E_\alpha\in\{1, e^{\pi i /3}, e^{2
  \pi i /3},-1, e^{4 \pi i /3}, e^{5 \pi i /3}\}E_\alpha \; .
\end{eqnarray}

\section{Conventions}
\subsection{Dirac Algebra}
Using $\sigma^0=\overline\sigma^0=1$, $\sigma^i = -
\overline{\sigma}^i$ and the 5D Dirac algebra
\begin{equation}
\gamma^\mu = \left(\begin{array}{cc} 0 & \sigma^\mu \\ \overline\sigma^\mu &0 \end{array} \right),\quad \gamma^5 = \left(\begin{array}{cc} -i & 0 \\ 0 &i \end{array} \right)
\end{equation}
we define the 6D Dirac algebra as
\begin{equation}
\Gamma^\mu = \left(\begin{array}{cc} 0 & \gamma^\mu \\ \gamma^\mu &0 \end{array} \right), \quad \Gamma^5 = \left(\begin{array}{cc} 0 & \gamma^5 \\ \gamma^5 &0 \end{array} \right),\quad
\Gamma^6 = \left(\begin{array}{cc} 0& -1 \\ 1 & 0\end{array} \right)
\end{equation}
and the 6D chirality operator as
\begin{equation}
i \Gamma^7 =  \left(\begin{array}{cc} -1 & 0 \\ 0 &1 \end{array} \right)
\end{equation}
In this basis,
\begin{equation}
\exp\left[\frac{\phi}{4}\left[\Gamma^5, \Gamma^6 \right]\right]=\mbox{diag}\left(e^{-i \phi/2},e^{i \phi/2},e^{i \phi/2},e^{-i \phi/2}\right)\sim U(1)\subset SO(1,5)
\end{equation}
corresponds to a counter-clockwise rotation with angle $\phi$ about the origin in the extra dimensional space,
$$\overline\Psi (\Gamma^5+i \Gamma^6)\Psi \longrightarrow e^{i \phi}\, \overline\Psi (\Gamma^5+i \Gamma^6)\Psi. $$
\subsection{Lie Algebras}
There are several ways to specify a weight vector \cite{Slansky:1981yr}. Let 
\begin{equation}
(\alpha^a)_k
\end{equation}
($a=1, \ldots, 6$) be the component vectors of the simple roots of $E_6$ and
\begin{equation}
A^{ij}=\frac{2 \alpha^i\cdot \alpha^j}{|\alpha^j|^2}
\end{equation} 
the (symmetric) Cartan Matrix of $E_6$. Then, there are
\begin{enumerate}
\item The weight vector in the canonical basis, $\mu_k=c^i(\alpha^i)_k$
\item The Dynkin coefficients $\Delta$ where $\Delta^a (A^{-1})^{ab} (\alpha^b)_i=\mu_i $
\item The dual weights $\overline{\mu}$ where $\Delta^i= A^{ji} \frac{2}{|\alpha^j|^2} \overline{\mu}^j $
\end{enumerate}
These are defined such that the scalar product of two weights in the canonical basis can be recast as
\begin{equation}
\mu \cdot \lambda = \Delta_{\mu} \cdot \overline{\lambda}=\Delta_{\lambda} \cdot \overline{\mu}
\end{equation}
In this paper we give shifts and charges as Dynkin coefficients and in the dual basis. 
\subsection{$E_6$ Representations}
The Tables \ref{tab27} and \ref{tab78} list the weights(roots) of the 
$\mathbf{27}$ and $\mathbf{78}$ representation of $E_6$ and the particle
assignments as used in this paper. The basis choice is not unique. We
define the Cartan generators in the dual basis as follows: 
\begin{eqnarray}
  \overline{I}^3_L&=& (-\frac{1}{2},-\frac{1}{2},-\frac{1}{2},-\frac{1}{2},-\frac{1}{2},\phantom{-}0) \\
  \overline{I}^{3c}_R&=& (\phantom{-}0,\phantom{-}0,-\frac{1}{2},\phantom{-}0,\phantom{-}0,\phantom{-}0) \\
  \overline{Q}_e&=& (-\frac{1}{3},-\frac{2}{3},-\frac{4}{3},-1,-\frac{2}{3},\phantom{-}0) \\
  \overline{Q}_Y&=& (\phantom{-}\frac{1}{6},-\frac{1}{6},-\frac{5}{6},-\frac{1}{2},-\frac{1}{6},\phantom{-}0) \\
  \overline{Q}_{B-L}&=&
  (\phantom{-}\frac{1}{6},-\frac{1}{6},-\frac{1}{3},-\frac{1}{2},-\frac{1}{6},\phantom{-}0)
  \\ 
  \overline{Q}_\chi&=&
  (\phantom{-}\frac{1}{2},-\frac{1}{2},\phantom{-}0,\phantom{-}\frac{1}{2},-\frac{1}{2},\phantom{-}0)
\end{eqnarray}
These vectors must be multiplied with the Dynkin coefficients of a state $|\mu\rangle$ to obtain the corresponding physical charges,
$$\overline{Q}_i \cdot \Delta(\mu)  = Q_i(\mu) $$
In this normalization,
\begin{eqnarray}
\overline Q_Y = \overline Q_{B-L} + \overline I^{3c}_R, \quad \overline Q_e = \overline Q_Y + \overline I^{3}_L 
\end{eqnarray}
\begin{table}
\begin{equation*}
\begin{small}
\begin{array}{l||rrrrrr||r|r|r|r|r|r||l}
\#&\lefteqn{\mbox{\bf $\Delta$ weight}}& & & & & & I_L & I_R^c & Q_e & Q_{Y}& Q_{B-L}& Q_\chi& \mbox{\bf assignment}  \\ \hline 
 1 & 1 & 0 & 0 & 0 & 0 & 0 & -\frac{1}{2} & 0 & -\frac{1}{3} & \frac{1}{6} &
   \frac{1}{6} & \frac{1}{2} & d_L \\ \hline
 2 & -1 & 1 & 0 & 0 & 0 & 0 & 0 & 0 & -\frac{1}{3} & -\frac{1}{3} &
   -\frac{1}{3} & -1 & D \\ \hline
 3 & 0 & -1 & 1 & 0 & 0 & 0 & 0 & -\frac{1}{2} & -\frac{2}{3} & -\frac{2}{3}
   & -\frac{1}{6} & \frac{1}{2} & u_R^c \\ \hline
 4 & 0 & 0 & -1 & 1 & 0 & 1 & 0 & \frac{1}{2} & \frac{1}{3} & \frac{1}{3} &
   -\frac{1}{6} & \frac{1}{2} & d_R^c \\ \hline
 5 & 0 & 0 & 0 & 1 & 0 & -1 & -\frac{1}{2} & 0 & -1 & -\frac{1}{2} &
   -\frac{1}{2} & \frac{1}{2} & e_L \\ \hline
 6 & 0 & 0 & 0 & -1 & 1 & 1 & 0 & 0 & \frac{1}{3} & \frac{1}{3} &
   \frac{1}{3} & -1 & D^c \\ \hline
 7 & 0 & 0 & 1 & -1 & 1 & -1 & -\frac{1}{2} & -\frac{1}{2} & -1 &
   -\frac{1}{2} & 0 & -1 & H_d \\ \hline
 8 & 0 & 0 & 0 & 0 & -1 & 1 & \frac{1}{2} & 0 & \frac{2}{3} & \frac{1}{6} &
   \frac{1}{6} & \frac{1}{2} & u_L \\ \hline
 9 & 0 & 0 & 1 & 0 & -1 & -1 & 0 & -\frac{1}{2} & -\frac{2}{3} &
   -\frac{2}{3} & -\frac{1}{6} & \frac{1}{2} & u_R^c \\ \hline
 10 & 0 & 1 & -1 & 0 & 1 & 0 & -\frac{1}{2} & \frac{1}{2} & 0 & \frac{1}{2}
   & 0 & -1 & H_u \\ \hline
 11 & 0 & 1 & -1 & 1 & -1 & 0 & 0 & \frac{1}{2} & \frac{1}{3} & \frac{1}{3}
   & -\frac{1}{6} & \frac{1}{2} & d_R^c \\ \hline
 12 & 1 & -1 & 0 & 0 & 1 & 0 & -\frac{1}{2} & 0 & -\frac{1}{3} & \frac{1}{6}
   & \frac{1}{6} & \frac{1}{2} & d_L \\ \hline
 13 & 0 & 1 & 0 & -1 & 0 & 0 & 0 & 0 & \frac{1}{3} & \frac{1}{3} &
   \frac{1}{3} & -1 & D^c \\ \hline
 14 & 1 & -1 & 0 & 1 & -1 & 0 & 0 & 0 & 0 & 0 & 0 & 2 & S \\ \hline
 15 & -1 & 0 & 0 & 0 & 1 & 0 & 0 & 0 & -\frac{1}{3} & -\frac{1}{3} &
   -\frac{1}{3} & -1 & D \\ \hline
 16 & 1 & -1 & 1 & -1 & 0 & 0 & 0 & -\frac{1}{2} & 0 & 0 & \frac{1}{2} &
   \frac{1}{2} & \nu_R^c \\ \hline
 17 & -1 & 0 & 0 & 1 & -1 & 0 & \frac{1}{2} & 0 & 0 & -\frac{1}{2} &
   -\frac{1}{2} & \frac{1}{2} & \nu_L \\ \hline
 18 & 1 & 0 & -1 & 0 & 0 & 1 & 0 & \frac{1}{2} & 1 & 1 & \frac{1}{2} &
   \frac{1}{2} & e_R^c \\ \hline
 19 & -1 & 0 & 1 & -1 & 0 & 0 & \frac{1}{2} & -\frac{1}{2} & 0 &
   -\frac{1}{2} & 0 & -1 & H_d \\ \hline
 20 & 1 & 0 & 0 & 0 & 0 & -1 & -\frac{1}{2} & 0 & -\frac{1}{3} & \frac{1}{6}
   & \frac{1}{6} & \frac{1}{2} & d_L \\ \hline
 21 & -1 & 1 & -1 & 0 & 0 & 1 & \frac{1}{2} & \frac{1}{2} & 1 & \frac{1}{2}
   & 0 & -1 & H_u \\ \hline
 22 & -1 & 1 & 0 & 0 & 0 & -1 & 0 & 0 & -\frac{1}{3} & -\frac{1}{3} &
   -\frac{1}{3} & -1 & D \\ \hline
 23 & 0 & -1 & 0 & 0 & 0 & 1 & \frac{1}{2} & 0 & \frac{2}{3} & \frac{1}{6} &
   \frac{1}{6} & \frac{1}{2} & u_L \\ \hline
 24 & 0 & -1 & 1 & 0 & 0 & -1 & 0 & -\frac{1}{2} & -\frac{2}{3} &
   -\frac{2}{3} & -\frac{1}{6} & \frac{1}{2} & u_R^c \\ \hline
 25 & 0 & 0 & -1 & 1 & 0 & 0 & 0 & \frac{1}{2} & \frac{1}{3} & \frac{1}{3} &
   -\frac{1}{6} & \frac{1}{2} & d_R^c \\ \hline
 26 & 0 & 0 & 0 & -1 & 1 & 0 & 0 & 0 & \frac{1}{3} & \frac{1}{3} &
   \frac{1}{3} & -1 & D^c \\ \hline
 27 & 0 & 0 & 0 & 0 & -1 & 0 & \frac{1}{2} & 0 & \frac{2}{3} & \frac{1}{6} &
   \frac{1}{6} & \frac{1}{2} & u_L
\end{array}
\end{small}
\end{equation*}
\caption{The Dynkin coefficients and particle assignments of the weights in the fundamental representation $\bf 27$ of $E_6$ and their quantum numbers. \label{tab27}}
\end{table}
\begin{table}
\begin{small}
\begin{eqnarray*}
\begin{array}{lr||rrrrrr||r|r|r|r|r|r||l}
\#& &\lefteqn{\mbox{\bf $\Delta$ root}}& & & & & & I_L & I_R^c & Q_e & Q_{Y}& Q_{B-L}& Q_\chi& \mbox{\bf assignment}  \\ \hline 
1 &  & 0&0&0&0&0&1 & 0 & 0 & 0 & 0 & 0 & 0 & \bf G \\ \hline
 2 &  & 0&0&1&0&0&-1 & -\frac{1}{2} & -\frac{1}{2} & -\frac{4}{3} &
   -\frac{5}{6} & -\frac{1}{3} & 0 & X \\ \hline
 3 &  & 0&1&-1&1&0&0 & -\frac{1}{2} & \frac{1}{2} & -\frac{1}{3} &
   \frac{1}{6} & -\frac{1}{3} & 0 & Q_{45} \\ \hline
 4 &  & 0&1&0&-1&1&0 & -\frac{1}{2} & 0 & -\frac{1}{3} & \frac{1}{6} &
   \frac{1}{6} & -\frac{3}{2} & Q_{16} \\ \hline
 5 &  & 1&-1&0&1&0&0 & -\frac{1}{2} & 0 & -\frac{2}{3} & -\frac{1}{6} &
   -\frac{1}{6} & \frac{3}{2} & Q_{16} \\ \hline
 6 &  & 0&1&0&0&-1&0 & 0 & 0 & 0 & 0 & 0 & 0 & \bf G \\ \hline
 7 &  & 1&-1&1&-1&1&0 & -\frac{1}{2} & -\frac{1}{2} & -\frac{2}{3} &
   -\frac{1}{6} & \frac{1}{3} & 0 & Q_{45} \\ \hline
 8 &  & -1&0&0&1&0&0 & 0 & 0 & -\frac{2}{3} & -\frac{2}{3} &
   -\frac{2}{3} & 0 & U_{45} \\ \hline
 9 &  & 1&-1&1&0&-1&0 & 0 & -\frac{1}{2} & -\frac{1}{3} & -\frac{1}{3}
   & \frac{1}{6} & \frac{3}{2} & D_{16} \\ \hline
 10 &  & 1&0&-1&0&1&1 & -\frac{1}{2} & \frac{1}{2} & \frac{1}{3} &
   \frac{5}{6} & \frac{1}{3} & 0 & X \\ \hline
 11 &  & -1&0&1&-1&1&0 & 0 & -\frac{1}{2} & -\frac{2}{3} & -\frac{2}{3}
   & -\frac{1}{6} & -\frac{3}{2} & U_{16} \\ \hline
 12 & & 1&0&0&0&1&-1 & -1 & 0 & -1 & 0 & 0 & 0 & \bf W_L^- \\ \hline
 13 &  & 1&0&-1&1&-1&1 & 0 & \frac{1}{2} & \frac{2}{3} & \frac{2}{3} &
   \frac{1}{6} & \frac{3}{2} & U_{16} \\ \hline
 14 &  & -1&0&1&0&-1&0 & \frac{1}{2} & -\frac{1}{2} & -\frac{1}{3} &
   -\frac{5}{6} & -\frac{1}{3} & 0 & X \\ \hline
 15 &  & -1&1&-1&0&1&1 & 0 & \frac{1}{2} & \frac{1}{3} & \frac{1}{3} &
   -\frac{1}{6} & -\frac{3}{2} & D_{16} \\ \hline
 16 &  & 1&0&0&1&-1&-1 & -\frac{1}{2} & 0 & -\frac{2}{3} & -\frac{1}{6}
   & -\frac{1}{6} & \frac{3}{2} & Q_{16} \\ \hline
 17 &  & 1&0&0&-1&0&1 & 0 & 0 & \frac{2}{3} & \frac{2}{3} &
   \frac{2}{3} & 0 & U_{45} \\ \hline
 18 &  & -1&1&0&0&1&-1 & -\frac{1}{2} & 0 & -1 & -\frac{1}{2} &
   -\frac{1}{2} & -\frac{3}{2} & L_{16} \\ \hline
 19 &  & -1&1&-1&1&-1&1 & \frac{1}{2} & \frac{1}{2} & \frac{2}{3} &
   \frac{1}{6} & -\frac{1}{3} & 0 & Q_{45} \\ \hline
 20 & & 0&-1&0&0&1&1 & 0 & 0 & 0 & 0 & 0 & 0 & \bf G \\ \hline
 21 &  & 1&0&1&-1&0&-1 & -\frac{1}{2} & -\frac{1}{2} & -\frac{2}{3} &
   -\frac{1}{6} & \frac{1}{3} & 0 & Q_{45} \\ \hline
 22 &  & -1&1&0&1&-1&-1 & 0 & 0 & -\frac{2}{3} & -\frac{2}{3} &
   -\frac{2}{3} & 0 & U_{45} \\ \hline
 23 &  & -1&1&0&-1&0&1 & \frac{1}{2} & 0 & \frac{2}{3} & \frac{1}{6} &
   \frac{1}{6} & -\frac{3}{2} & Q_{16} \\ \hline
 24 &  & 0&-1&1&0&1&-1 & -\frac{1}{2} & -\frac{1}{2} & -\frac{4}{3} &
   -\frac{5}{6} & -\frac{1}{3} & 0 & X \\ \hline
 25 &  & 0&-1&0&1&-1&1 & \frac{1}{2} & 0 & \frac{1}{3} & -\frac{1}{6} &
   -\frac{1}{6} & \frac{3}{2} & Q_{16} \\ \hline
 26 &  & 1&1&-1&0&0&0 & -\frac{1}{2} & \frac{1}{2} & \frac{1}{3} &
   \frac{5}{6} & \frac{1}{3} & 0 & X \\ \hline
 27 &  & -1&1&1&-1&0&-1 & 0 & -\frac{1}{2} & -\frac{2}{3} &
   -\frac{2}{3} & -\frac{1}{6} & -\frac{3}{2} & U_{16} \\ \hline
 28 &  & 0&-1&1&1&-1&-1 & 0 & -\frac{1}{2} & -1 & -1 & -\frac{1}{2} &
   \frac{3}{2} & E_{16} \\ \hline
 29 &  & 0&-1&1&-1&0&1 & \frac{1}{2} & -\frac{1}{2} & \frac{1}{3} &
   -\frac{1}{6} & \frac{1}{3} & 0 & Q_{45} \\ \hline
 30 &  & 0&0&-1&1&1&0 & -\frac{1}{2} & \frac{1}{2} & -\frac{1}{3} &
   \frac{1}{6} & -\frac{1}{3} & 0 & Q_{45} \\ \hline
 31 &  & 2&-1&0&0&0&0 & -\frac{1}{2} & 0 & 0 & \frac{1}{2} &
   \frac{1}{2} & \frac{3}{2} & L_{16} \\ \hline
 32 &  & -1&2&-1&0&0&0 & 0 & \frac{1}{2} & \frac{1}{3} & \frac{1}{3} &
   -\frac{1}{6} & -\frac{3}{2} & D_{16} \\ \hline
  33 & & 0& -1&2&-1&0&-1 & 0 & -1 & -1 & -1 & 0 & 0 & E_{45}\equiv \bf W_R^- \\ \hline
 34 &  & 0&0&-1&2&-1&0 & 0 & \frac{1}{2} & 0 & 0 & -\frac{1}{2} &
   \frac{3}{2} & \nu_{16} \\ \hline
 35 &  & 0&0&0&-1&2&0 & -\frac{1}{2} & 0 & -\frac{1}{3} & \frac{1}{6}
   & \frac{1}{6} & -\frac{3}{2} & Q_{16} \\ \hline
 36 &  & 0&0&-1&0&0&2 & \frac{1}{2} & \frac{1}{2} & \frac{4}{3} &
   \frac{5}{6} & \frac{1}{3} & 0 & X
\end{array}
\end{eqnarray*}
\end{small}
\caption{The positive roots of $E_6$\label{tab78}, their Dynkin
  coefficients and quantum numbers. The naming scheme is with respect
  to the $E_6\nobreak\rightarrow\nobreak SO(10)\times U(1)_\chi$
  branching $\bf 78 \rightarrow 45_0 + 16_{-3/2} + \overline{16}_{3/2}
  + 1_0$ as it is relevant to this paper. The six zero roots are not
  shown, the elements of the $G_{LR}$ algebra are highlighted.} 
\end{table}
\clearpage
\section{Matter and Gauge Multiplets on the $T^2/\mathbb Z_n$ orbifold}
\label{apx:orbisusy}
\subsection{Gauge Theory}
Upon compactification, the 6D Lorentz group is broken to \begin{equation} SO(1,5) \rightarrow SO(1,3)\times U(1)\end{equation} 
so the 6D vector is split into a 4D vector $A_\mu$ and a scalar in the adjoint of the gauge group, \begin{equation}\Sigma=\frac{1}{\sqrt{2}}(A_6+ i A_5).\end{equation}
The latter transform nontrivially under the $\mathbb Z_n \subset U(1)$ of the orbifold space group. In particular, under $\mathbb Z_n$ rotations $x_5-i x_6 \rightarrow e^{-2 \pi i /n} (x_5-i x_6)$,
\begin{equation}
\Sigma \rightarrow e^{-2\pi i /n} \Sigma
\end{equation}
If there is an additional gauge shift associated with the orbifold action, the vector components belonging to a root $E_\alpha$ transform as
\begin{equation}\label{eq:orbitra}
A^\alpha_\mu \rightarrow e^{2 \pi i V\cdot \alpha} A^\alpha_\mu, \quad \Sigma^\alpha \rightarrow e^{2 \pi i V\cdot \alpha} e^{- 2\pi i/n} \Sigma^\alpha
\end{equation}
Massless modes have constant Kaluza-Klein wavefunctions in the extra dimension, 
and therefore only appear for components which are invariant under $r_6$. This means that the
unbroken gauge group is determined by the set of roots $\alpha$ which satisfy $V \cdot \alpha \in \mathbb Z$, 
which naturally includes the zero weights and thus rank is preserved. The zero modes of scalars correspond to roots which are broken on all fixed points. 
In that sense the situation is similar to a $S^1/\mathbb Z_2$ orbifold 
where massless scalars appear if a generator is broken on both ends of the fundamental domain.

\subsection{Supersymmetric theory\label{sect:susy}}
Hypermultiplets coupled to supersymmetric Yang-Mills theory in 5 and 6 dimensions can be formulated using $\mathcal{N}=1$ superfields \cite{HEIDISUSY}. The field content is
identical in both cases, namely that of 4D $\mathcal{N}=2$ supersymmetry.
In this language, the gauge hypermultiplet consists of a vector superfield $\hat V$ and a chiral superfield $\hat \chi$ in the adjoint. The physical scalar component of the latter 
contains the extra dimensional gauge field components, $\hat \chi| =
\Sigma + \mathcal O(\theta)$. There are now three real auxiliary
fields forming a triplet under the R symmetry, $SU(2)^R$. 
The hypermultiplet can be written as two chiral superfields with opposite charge, $\hat \Phi$ and $\hat \Phi_c$.
The chiral fermionic SUSY parameters can be defined as a 6D spinor $\xi=(\xi_{1}, \overline{\xi}_2,0,0)^T$. In general, these parameters transform nontrivially
under orbifold rotations. 
Given the 6D Dirac matrices, $\Gamma^M$, the $\mathbb Z_n$ rotations in the 5-6
plane of orbifolds on $T^2$ are generated by $$ \theta =\exp\left[
\frac{2 \pi}{n}\frac{1}{4}[\Gamma^5,\Gamma^6]\right].$$ 
With trivial embedding in the R symmetry, the SUSY parameter transforms as
\begin{equation}\label{eq:spintra}
\left(\begin{array}{l} \xi_1 \\ \overline{\xi}_2
\end{array}\right) \stackrel{\theta}{\longrightarrow} \left(\begin{array}{l}
  e^{-i \pi /n}\xi_1 \\ e^{i  \pi /n }\overline{\xi}_2
 \end{array}\right) 
\end{equation}
which would result in a non-supersymmetric massless spectrum, and $\theta^n\neq 1$.

It is known from 10D orbifold constructions (such
as heterotic models), that certain conditions have to be fulfilled in order to
preserve at least $\mathcal N=1$ SUSY in 4D\footnote{We
thank P. Vaudrevange for clarifications concerning this problem.}. 
The smallest
nontrivial spinor in 10D is both  chiral and real, and thus
corresponds to 4D $\mathcal N=4$ SUSY. General $\mathbb Z_n$
transformations on the complex torus coordinates are generated by   
\begin{equation}
  (z_A,z_B,z_C) \rightarrow (e^{i A} z_A,e^{i B} z_B,e^{i C} z_C) \, .
\end{equation}
Using a 10D dirac algebra, the spinors then transform as 
\begin{equation}
\label{general10dzn}
\theta = \exp\left[\frac{A}{4} [\Gamma^5,\Gamma^6] +\frac{B}{4}
  [\Gamma^7,\Gamma^8] +\frac{C}{4} [\Gamma^9,\Gamma^{10}]\right] 
\end{equation}
The condition that we want to preserve at least one 4D supersymmetry
can now be formulated as a constraint on the coefficients, 
\begin{equation}
\label{eqn1susy}
  A+B+C=0 \quad ,
\end{equation}
which is not possible in 6D, where $B=C \equiv 0$. However, after trivial dimensional reduction
to 6 dimensions, the $SO(4)\simeq SU(2)\times SU(2)'$ Lorentz algebra of the higher dimensions generated by $\Gamma^7\dots \Gamma^{10}$
becomes an internal symmetry. It contains the $SU(2)^R$
under which the chiral 6D $\mathcal{N}=1$ SUSY generator is
a doublet. Thus, we have the choice to either introduce additional (possibly small) spacetime
dimensions with nontrivial rotational phases that cancel the chiral phases to preserve 4D $\mathcal N=1$ supersymmetry (this however might be viable only in the context of a heterotic string theory). Or we can start with a 6D $\mathcal N=1$ setup and assign
an embedding of $\mathbb Z_n$ in $I^{3R}$ of $SU(2)^R$ to the chiral 6D spinors, $$\theta = 
\exp\left[\frac{2 \pi}{n}\frac{1}{4}([\Gamma^5,\Gamma^6] + c_R i I^{3R} )\right]  $$
where the constant $c_R$ can be chosen
appropriately to conserve at least one 4D supersymmetry, and $\theta^n=1$. 
From now on we assume that the conserved 4D $\mathcal{N}=1$ supersymmetry is chosen such that it acts as usual within the $\mathcal{N}=1$ superfields $\hat V,\hat \chi,\hat \Phi,\hat \Phi^c$.
For the orbifold construction this means that orbifold phases are assigned to entire superfields, and the resulting massless multiplets are given
directly as zero modes of complete $\mathcal{N}=1$ superfields. The transformations in (\ref{eq:orbitra}) are generalized to
\begin{equation}\label{eq:superorbitra}
\hat V^\alpha \rightarrow e^{2 \pi i V\cdot \alpha} \hat V^\alpha, \quad \hat \chi^\alpha \rightarrow e^{2 \pi i V\cdot \alpha} e^{- 2\pi i/n} \hat \chi^\alpha
\end{equation}
Then, zero modes of $A_\mu$ automatically come with complete 4D vector multiplets, while zero modes of $A_5$ and $A_6$ come with complete 4D chiral multiplets. The discussion of ordinary YM theory can therefore be generalized to the SYM case in a straightforward manner.

\end{document}